\documentclass[preprint,aps,12pt,notitlepage,nofootinbib,tightenlines]{revtex4}
\usepackage{amsmath}
\usepackage{enumerate}
\usepackage{bm}
\usepackage{times}
\usepackage{braket}
\usepackage{color}
\usepackage{epsfig}
\usepackage{slashed}
\usepackage{hyperref}
\usepackage{multirow}
\usepackage{booktabs}
\usepackage{array}
\usepackage{float}
\newcommand{\beq}{\begin{eqnarray}}
\newcommand{\eeq}{\end{eqnarray}}
\newcommand{\be}{\begin{equation}\begin{aligned}}
\newcommand{\ee}{\end{aligned}\end{equation}}

\newcommand{\gev}{\text{GeV}}

\definecolor{Red}{rgb}{1.,0.,0.}

\definecolor{Blue}{rgb}{0.,0.,1.}

\definecolor{nicered}{rgb}{0.7,0.1,0.1}
\definecolor{nicegreen}{rgb}{0.1,0.5,0.1}
\def\lsim{ {\ \lower-1.2pt\vbox{\hbox{\rlap{$<$}\lower6pt\vbox{\hbox{$\sim$}}}}\ } }
\def\gsim{ {\ \lower-1.2pt\vbox{\hbox{\rlap{$>$}\lower6pt\vbox{\hbox{$\sim$}}}}\ } }

\hypersetup{colorlinks,citecolor=nicegreen,linkcolor=nicered}
\begin{document}
\title{Searching for single production of vectorlike quarks decaying into $Wb$ \\ at a future muon-proton collider}
\author{Jin-Zhong Han$^{1}$\footnote{E-mail: hanjinzhong@zknu.edu.cn}}
\author{Yao-Bei Liu$^{2}$\footnote{E-mail: liuyaobei@hist.edu.cn}}
\author{Stefano Moretti$^{3,4}$\footnote{E-mail: stefano.moretti@cern.ch}}
\affiliation{1. School of Physics and
Telecommunications Engineering, Zhoukou Normal University, Zhoukou 466001, P.R. China\\
2. Henan Institute of Science and Technology, Xinxiang 453003, P.R. China \\  
3. School of Physics \& Astronomy, University of Southampton, Highfield, Southampton SO17 1BJ, UK \\
4. Department of Physics \& Astronomy, Uppsala University, Box 516, 751 20 Uppsala, Sweden}
\begin{abstract}
This work investigates the discovery potential for singly produced vectorlike quarks~(VLQs) $T$ ($Q=+2/3e$) and $Y$ ($Q=-4/3e$) decaying to $Wb$ at future $\mu p$ colliders with $\sqrt{s}=5.29$, 6.48, and 9.16 TeV, analyzing both leptonic and hadronic $W$ decay channels through detailed detector simulations. The hadronic channel demonstrates superior sensitivity, enabling $5\sigma$ discovery up to $m_T=3750$ GeV ($m_Y=4100$ GeV) at 9.16 TeV with 100 fb$^{-1}$, while exclusion limits reach $m_T=4500$ GeV ($g^*\geq0.06$) and $m_Y=4800$ GeV ($\kappa_Y\geq0.04$) - significantly beyond LHC capabilities. At 5.29 TeV, discovery regions cover $g^*\in[0.15,0.5]$ for $m_T\in[1500,2520]$ GeV and $\kappa_Y\in[0.12,0.5]$ for $m_Y\in[1700,2700]$ GeV, with exclusions extending to $m_T=2750$ GeV and $m_Y=3020$ GeV. These results, obtained within a simplified two-parameter framework ($m_{T/Y}$ and electroweak couplings $g^{\ast}$), establish $\mu p$ colliders as uniquely powerful tools for probing high-mass VLQ states, particularly in the boosted jet regime.
\end{abstract}

\maketitle

\newpage
\section{Introduction}
Vectorlike quarks (VLQs)  with masses at the TeV scale are generally  predicted  in  a  variety of extensions of the Standard Model~(SM), such as the little Higgs models~\cite{Arkani-Hamed:2002iiv,ArkaniHamed:2002qy,Han:2003wu,Chang:2003vs}, composite Higgs models~\cite{Agashe:2004rs,Contino:2006qr,Lodone:2008yy,Matsedonskyi:2012ym}, two-Higgs-doublet models~\cite{Benbrik:2022kpo,Arhrib:2024tzm,Arhrib:2024dou,Arhrib:2024nbj,Benbrik:2024hsf,Arhrib:2024mbq}, and other extended models~\cite{He:1999vp,Wang:2013jwa,He:2001fz,He:2014ora}. A common feature of these new particles is that the left- and right-handed chiral components  transform in the same way under the electroweak (EW) symmetry group of the SM~\cite{Aguilar-Saavedra:2013qpa}. Unlike for chiral quarks, bare mass terms of VLQs are gauge invariant and therefore they can avoid the strict constraints by the Higgs boson data\footnote{An extra fourth generation of SM-like quarks~\cite{He:2001tp,Chen:2012wz} should be much heavier due to the EW precision constraints.}. Moreover,  VLQs  have the potential to stabilize the EW vacuum~\cite{Xiao:2014kba,Cingiloglu:2023ylm}, the so-called Cabibbo-Kobayashi-Maskawa (CKM) unitarity problem~\cite{Cheung:2020vqm,Crivellin:2022rhw,Belfatto:2021jhf,Branco:2021vhs,Botella:2021uxz}, and may also provide explanations for various experimental anomalies, such as the
$W$-mass one~\cite{Cao:2022mif,Crivellin:2022fdf,He:2022zjz,Branco:2022gja,Abouabid:2023mbu}. There exists three types of multiple VLQs, including EW singlet ($T$, $ B$),  doublets [$\left(X,T\right),\left(T,B\right)$ or $\left(B,Y\right)$], and  triplets [$\left(X,T,B\right)$ or $\left(T,B,Y\right)$].
 In the  models embedding these, VLQs are expected to couple preferentially to third-generation  quarks and can  give rise to a rich variety of phenomena at the Large Hadron Collider (LHC)  and future high-energy colliders~(for example,  see~\cite{DeSimone:2012fs,Buchkremer:2013bha,Aguilar-Saavedra:2009xmz,Mrazek:2009yu,Dissertori:2010ug,Atre:2011ae,Cacciapaglia:2011fx,Cacciapaglia:2012dd,Gopalakrishna:2013hua,Matsedonskyi:2014mna,Backovic:2014uma,
Chen:2016yfv,Fuks:2016ftf,Liu:2016jho,Aguilar-Saavedra:2017giu,Cui:2022hjg,Xie:2019gya,Benbrik:2019zdp,Aguilar-Saavedra:2019ghg,Belyaev:2021zgq,
Bhardwaj:2022nko,Bhardwaj:2022wfz,Verma:2022nyd,Bardhan:2022sif,Alves:2023ufm,Canbay:2023vmj,Belyaev:2023yym,Han:2023ied,Liu:2024hvp,
Shang:2024wwy,Cetinkaya:2020yjf,Zhang:2024ncj,Yang:2024aav}).

The VLQ $T$-state~(VLQ-$T$, also denoted $T$ in the remainder) has an electric charge of $+2/3$e and and occurs in any weak-isospin multiplet, whereas the VLQ $Y$-state~(VLQ-$Y$, also denoted by $Y$ in the remainder), which has an electric charge of $-(4/3)$e, only appears in doublet or triplet representations. The singlet VLQ-$T$ has three possible decay modes: $T\to$  $bW$, $tZ$, and $th$ (unless an extended Higgs sector is present \cite{Aguilar-Saavedra:2017giu,Benbrik:2019zdp}). In the high-mass limit, the  branching ratios (BRs) are BR$(T\to th)\approx {\rm BR}(T\to tZ)\approx \frac{1}{2}{\rm BR}(T\to Wb)$. Due to its charge, the VLQ-$Y$ can decay only
into $Wb$ pairs with same charge with a BR of 100\%.
Up to now, the ATLAS and CMS Collaborations have conducted extensive VLQ searches for the (QCD induced) pair-production processes  and the constraints on their masses have been obtained at a 95\% confidence level~(CL)~\cite{ATLAS:2024gyc,ATLAS:2022hnn,CMS:2022fck,Aaboud:2018wxv,Aaboud:2018xpj,Aaboud:2018uek,Aaboud:2018ifs,Sirunyan:2018qau,Sirunyan:2018omb,Aaboud:2018pii,CMS:2019eqb,Buckley:2020wzk}. For instance, the minimum mass of a singlet VLQ-$T~(Y)$ is set at about 1.36~(1.7) TeV
from direct searches by the ATLAS Collaboration with an integrated luminosity of 140 fb$^{-1}$~\cite{ATLAS:2024gyc}. The CMS Collaboration have excluded a singlet~(doublet)  VLQ-$T$ mass below 1.46~(1.48) TeV at 95\% CL by using 138 fb$^{-1}$ of $pp$ collision data in the leptonic  final states~\cite{CMS:2022fck}.
 In addition, such VLQ-$T$ can be singly produced at the LHC via  EW  interactions and the corresponding processes are highly sensitive to the couplings between VLQs and SM quarks~\cite{Moretti:2016gkr,
Carvalho:2018jkq,Deandrea:2021vje}.  Searches performed recently by the ATLAS
and CMS Collaborations set limits on VLQs masses and couplings using Run 2 recorded
data~\cite{ATLAS:2022ozf,ATLAS:2023pja,ATLAS:2023bfh,ATLAS:2024xdc,ATLAS:2024kgp,CMS:2023agg,CMS:2024qdd,CMS:2024bni}.
For a benchmark signal prediction of a $SU(2)_L$
singlet VLQ-$T$ with the mixing parameter $\kappa=0.5$, which governs VLQ interactions with SM particles,  masses below 2.1 TeV are excluded by ATLAS with an integrated
luminosity of 139 fb$^{-1}$~\cite{ATLAS:2024xdc}. For the VLQ-$Y$, the strongest exclusion limit is set by ATLAS in the $bbbq'$ final state, where $\kappa\geq 0.3$ is excluded for mass near 2 TeV \cite{ATLAS:2024kgp}.

 In this work, we consider a muon-proton collider with
multi-TeV beam energies, which was proposed two decades ago~\cite{Shiltsev:1997pv,Ginzburg:1998yw,Cheung:1997rg,Cheung:1999wy,Carena:2000su},  and more
recently in Refs.~\cite{Kaya:2018kyt,Kaya:2019ecf,Ketenoglu:2022fzo,Dagli:2022idi,Kaya:2022wrc,Akturk:2024evo}. Compared to  a $ep$ collider, a $\mu p$ one can achieve a much higher center-of-mass energy and thus exotic particles production is  more probable due to much larger
scattering cross sections. Further,  beyond the SM~(BSM) studies at these types of machines
usually suffer from smaller QCD backgrounds than at $pp$ colliders. Recently, a lot of related
phenomenological work has been carried out for a future $\mu p$ collider~\cite{Caliskan:2017meb,Acar:2017eli,Caliskan:2018vep,Caliskan:2018,Alici:2019asv,Ozansoy:2019rmu,Spor:2020rig,Cheung:2021iev,Aydin:2021iky,Gurkanli:2024tfo,Alici:2024eez}.
Phenomenological studies of VLQs at an $ep$ colliders can be found in Refs.~\cite{Liu:2017rjw,Han:2017cvu,Zhang:2017nsn,Gong:2020ouh,Shang:2022tkr}.
In this study, we will focus on the observability of single $T/Y$
production at a future $\mu p$ collider via the $T/Y\to Wb$ decay channel both in the semileptonic and hadronic final states.

The paper is arranged as follows. In Sec. II, we consider a simplified model
including the VLQ-$T/Y$ and present the cross sections for the single production process at a $\mu p$ collider with three different center-of-mass energies. In Sec.~III, we discuss its observability via the decay modes $T/Y\to  bW\to b\ell+\slashed E_T$ and $T/Y\to bW\to b jj$, where $\ell$ is a lepton, $\slashed E_T$ is the missing transverse energy,  and $j$ is a jet (notice that the two jets will  merge into a single fat one $J$). Finally, conclusions are presented in Sec.~IV.

\section{The simplified model and single $T/Y$ production at a $\mu p$ collider}
Following the notation of Ref.~\cite{Buchkremer:2013bha}, a generic parametrization of an effective Lagrangian for a singlet VLQ-$T$ is given by
\beq
\hspace*{-0.15cm}
{\cal L}_{\rm T} =\frac{gg^{\ast}}{2}\{\frac{1}{\sqrt{2}}[\bar{T}_{L}W_{\mu}^{+}
    \gamma^{\mu} b_{L}]+
    \frac{1}{2c_W}[\bar{T}_{L} Z_{\mu} \gamma^{\mu} t_{L}]
    - \frac{m_{T}}{2m_{W}}[\bar{T}_{R}ht_{L}] -\frac{m_{t}}{2m_{W}} [\bar{T}_{L}ht_{R}]\}+ H.c.,
  \label{TsingletVL}
\eeq
where $g$ is the $SU(2)_L$ gauge coupling constant, and $\theta_W$ is the Weinberg angle.  Thus, there are only two model parameters: the VLQ-$T$ quark mass $m_T$ and the coupling strength to SM quarks in units of standard couplings, $g^{\ast}$.
As mentioned, the singlet VLQ-$T$ has three possible decay modes: $T\to$  $bW$, $tZ$, and $th$. For $M_T\geq 1~$TeV, the  BRs  are BR$(T\to th)\approx {\rm BR}(T\to tZ)\approx \frac{1}{2}{\rm BR}(T\to Wb)$, which is a good approximation as expected from the Goldstone
boson equivalence theorem~\cite{He:1992nga,He:1993yd,He:1994br,He:1996rb,He:1996cm}.

 Certainly, the coupling parameter, $g^{\ast}$, can also be described as other constants, i.e., $s_{L}$ in ~\cite{Aguilar-Saavedra:2013qpa} and $\kappa_{T}$~\cite{Buchkremer:2013bha}. A simple relationship among these coupling parameters is:
$g^{\ast}=\sqrt{2}\kappa_{T}=2s_{L}$~\cite{Benbrik:2024fku}. For $m_{T}= 1.5~(1.8)$~TeV, $g^{\ast}$ should be smaller than about 0.42~(0.56)~\cite{Benbrik:2024fku}.
 In this work we take only a phenomenologically guided limit: $g^{\ast}\leq 0.5$, in the region $m_{T}\geq 1.5$~TeV.

  Assuming that the VLQ-$Y$ only couples to the
SM third generation quarks, an effective Lagrangian framework can be written as
\beq
{\cal L}_{\rm Y} =&& \frac{g\kappa_{Y}^{L/R}}{\sqrt{2}}[\bar{Y}_{L/R}W_{\mu}^{-}
    \gamma^{\mu} b_{L/R} ]+ H.c.
  \label{VLQ-Y}
\eeq
Note that we can assume $\kappa_{Y}^{L}=0$ for a $\left(B,Y\right)$ doublet, and
$\kappa_{Y}^{R}=0$ for a $\left(T, B,Y\right)$ triplet~\cite{Aguilar-Saavedra:2013qpa}.
The cross sections for the single production process and the kinematics of the final states studied here are similar for left-handed and right-handed couplings.
For simplicity, we consider a benchmark scenario with right-handed couplings only in this work:
$\kappa_{Y}=\kappa_{Y}^{R}\neq 0$ and $\kappa_{Y}^{L}=0$, as for example in the case of the ($B,Y$) doublet.
For large mass values of the VLQ-$Y$, the decay width is approximated as $\Gamma_{Y}\sim \kappa_{Y}^{2}m_{Y}^{3}\times 3.28\times10^{-7}$~GeV$^{-2}$~\cite{Cetinkaya:2020yjf}.

\begin{figure}[h]
\centering
\includegraphics[width = 16cm ]{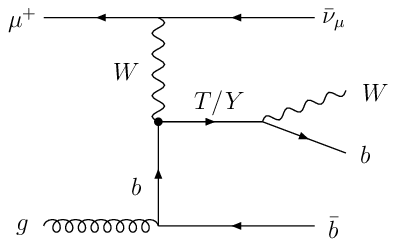}
\vspace{-18cm}
\caption{Representative tree-level Feynman diagram for single production of VLQ-$T/Y$ at a $\mu p$ collider followed by the $T/Y\to bW$ decay channel.}
\label{fig:fey}
\end{figure}

From the above discussions, we know that the VLQ-$T/Y$  can be singly produced in $\mu p$ collisions via
$Wb$ fusion process with a subsequent decay $T/Y\to Wb$. An example of a leading order (LO) Feynman diagram is depicted in Fig.~\ref{fig:fey}.

The LO cross sections are obtained using MadGraph5-aMC$@$NLO~\cite{Alwall:2014hca} with default NNPDF23L01 parton distribution functions (PDFs)~\cite{Ball:2014uwa} taking the default renormalization and factorization scales. The beam energies are  taken as three typical center-of-mass~(c.m.) energies:
  \begin{itemize}
\item
$\sqrt{s}= 5.29$~TeV, with $E_p=7$ TeV, and $E_\mu= 1.0$~TeV,
\item
$\sqrt{s}= 6.48$~TeV, with $E_p=7$ TeV, and $E_\mu= 1.5$~TeV,
\item
$\sqrt{s}= 9.16$~TeV, with $E_p=7$ TeV, and $E_\mu= 3.0$~TeV,
\end{itemize}
respectively.

\begin{figure}[thb]
\begin{center}
\vspace{-0.5cm}
\centerline{\epsfxsize=9cm \epsffile{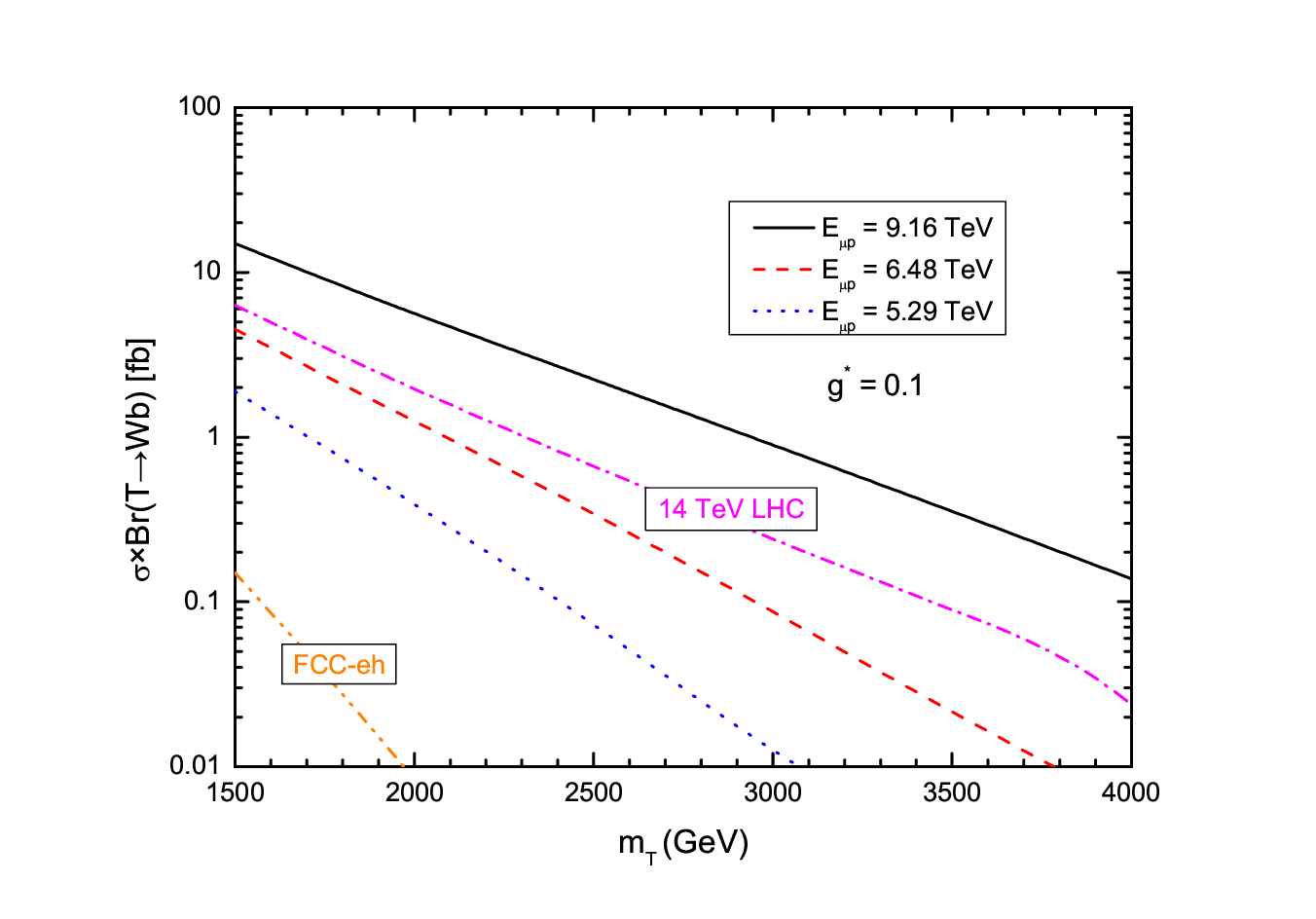}\epsfxsize=9cm \epsffile{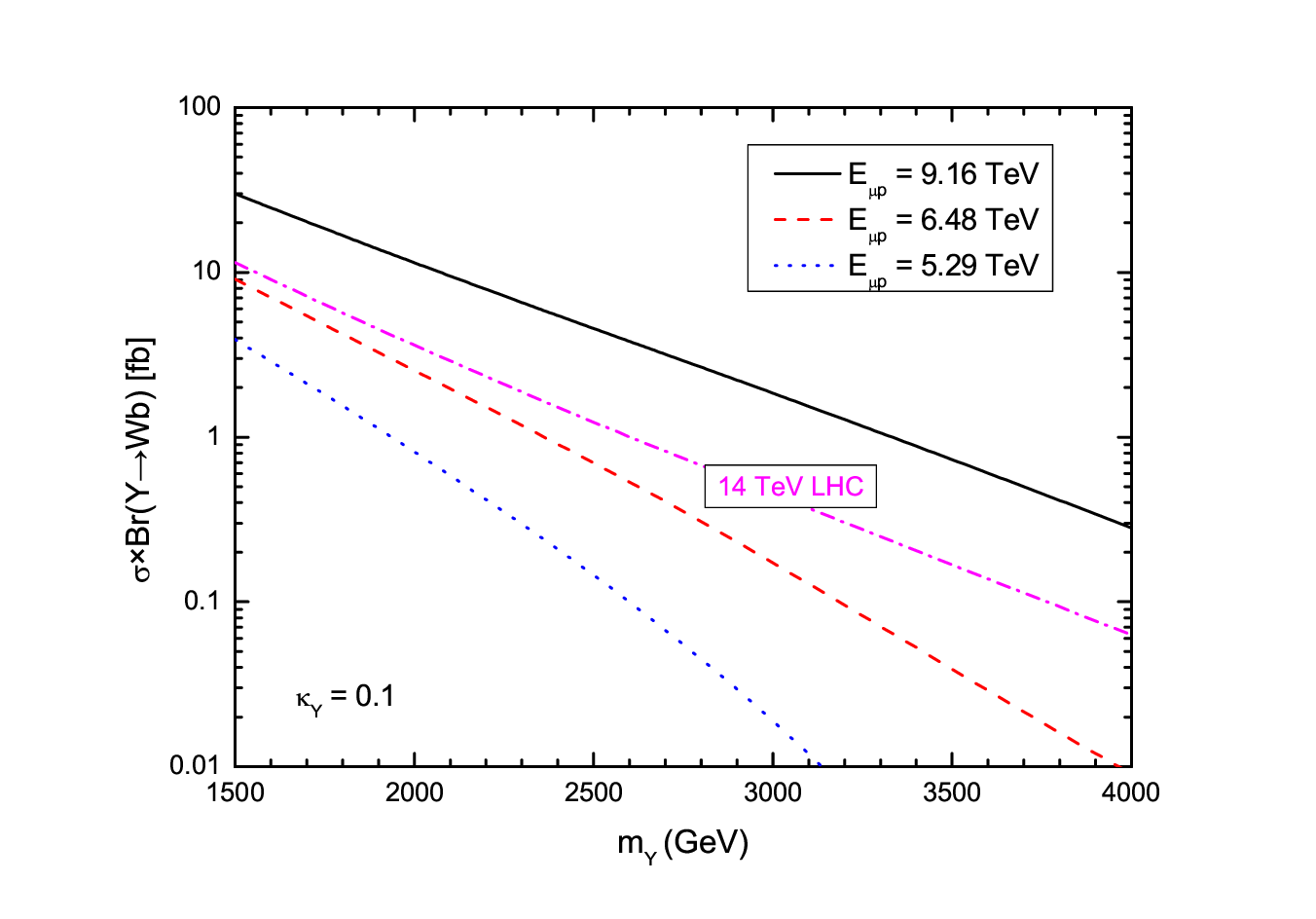}}
\caption{Cross section of $\sigma\times{\rm BR}(T/Y\to bW)$ as a function of the VLQ mass for $g^{*}/\kappa_{Y}=0.1$  at a $\mu p$ collider with three different c.m. energies. (The LHC and FCC-eh results for competing processes are also shown: see the text for details.)}
\label{cross}
\end{center}
\end{figure}

In Fig.~\ref{cross}, we show the dependence of the cross sections  $\sigma\times{\rm BR}(T/Y\to bW)$  on the VLQ-$T/Y$ mass for $g^{*}/\kappa_{Y}=0.1$. Note that  the conjugate processes have also been included~\footnote{All calculations assume unpolarized initial states, providing conservative estimates that facilitate direct comparison with existing collider results.}.
As the VLQ mass grows, the cross section of single production decreases slowly due to a smaller phase space.
For comparison, we also display the cross sections for the single VLQ-$T$ production processes at the 14 TeV LHC, $\sigma(pp\to Tbj)\times{\rm BR}(T\to bW)$ , and the
electron-hadron Future Circular Collider (FCC-eh), $\sigma(ep\to \nu T j)\times{\rm BR}(T(\to bW))$ with $\sqrt{s}= 3.46$ TeV, respectively. We also find that the cross sections can be up
to about one order of magnitude larger than those at the
FCC-eh, and even comparable to those at the
LHC with $\sqrt{s}=14$ TeV. For $g^{\ast}/\kappa_{Y}=0.1$ and $m_{T/Y}=2$ TeV, the cross section can reach 0.4~(0.82), 1.3~(2.54), and 5.6~(11.4) fb, respectively, at a $\mu p$ collider with three different c.m. energies. Certainly, the single production cross section is proportional to the square of the coupling strength $g^{\ast}$ or $\kappa_{Y}$.

\section{Collider simulation and analysis}
\begin{figure}[htb]\vspace{0.5cm}
\begin{center}
\centerline{\epsfxsize=12cm \epsffile{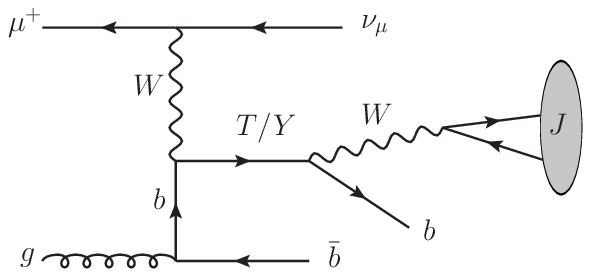}}
\vspace{-12cm}
\caption{ Representative Feynman diagram for the $b+J$ final state.}
\label{fatjet}
\end{center}
\end{figure}

Next we analyze the observation potential of the discussed $\mu p$ processes by performing a
Monte Carlo (MC) simulation of  signals (and, eventually, also background) events generated by single VLQ-$T/Y$ production and the $T/Y\to bW$ decay channel.  Considering the case of the  $W$ boson decaying leptonically, the final events are required to have exactly one isolated identified lepton~(specifically, an electron),  at least one $b$-tagged jet and large missing transverse energy from the escaping neutrino. (We do not consider the selection for the associated bottom quark from the splitting of an initial-stage gluon into a pair of $b$-quarks, which is often outside of the detector acceptance due to its typically low momentum.)
To increase the signal rate, we also impose that the $W$ boson decays into a  pair of quarks where the emerging jets $j$ can be collimated so as to appear as a fat jet~($J$), for which a representative Feynman diagram is given in Fig.~\ref{fatjet}.

Note that the kinematics of the final states are similar for VLQ-$T$ and VLQ-$Y$, so that the acceptances for the two types of VLQs
are found to be the same. Thus, the VLQ-$Y$ signals were not simulated separately. Finally, since we are including charged conjugation and the charge of the jets is not reconstructed, the contributions to the signal
of the $T$ and $Y$ mediated processes are indistinguishable so that, eventually, they can be summed over despite, in the remainder, they
are considered separately (thus, by implicitly exploiting MC truth knowledge).

\subsection{Expected discovery and exclusion reach for VLQ-$T$}
Next we perform a the signal-to-background analysis for the above two final states, which signal patterns are (to recap)
\begin{enumerate}[(i)]
 \item
$\mu^{+}p\to \bar{\nu}_{\mu}T(\to bW^{+})\bar{b}\to  e^{+}b\bar{b}+\slashed E_{T}$ for $W^{+}\to \ell^{+}\nu_{\ell}$ with $\ell= e$.
\item
$\mu^{+}p\to \bar{\nu}_{\mu}T(\to bW^{+})\bar{b}\to  J+b\bar{b}+\slashed E_{T}$ for $W^{+} \to q\bar{q}'$ with $q,q' \in \{u,d,s,c\}$.
\end{enumerate}

Note that, for the leptonic $W$ decay~($W_{\text{lep}}$) mode, we consider the final leptons to be only electrons due to large SM background processes $\mu p\to \mu jj$ and $\mu p\to\mu b\bar{b}$ whereas, for the hadronic decay~($W_{\text{had}}$) mode, the $W$ boson is boosted and thus its decay
products have low angular separation and are reconstructed as a fat one $J$ (as intimated).  (In
our simulation, the conjugate processes of all  backgrounds
 have also been considered.) The dominant SM backgrounds for the first final states come from the following processes:
 \begin{itemize}
 \item $\mu p\to \nu_{\mu}tb$ with $t\to bW\to be\nu_{e}$;
\item $\mu p\to  \nu_{\mu}Wj$ with $W\to e\nu_{e}$.
\end{itemize}
For the second final states including fat jet, the dominant SM backgrounds come from the following processes:
\begin{enumerate}[(i)]
  \item $\mu p\to \nu_{\mu}tb$ with $t\to bW\to bjj$;
\item $\mu p\to  \nu_{\mu}Wj$ with $W\to jj$;
\item $\mu p\to  \nu_{\mu}Zj$ with $Z\to q\bar{q}$  and $Z\to b\bar{b}$;
\item $\mu p\to \nu_{\mu}jj$.
\end{enumerate}

The presented cross sections include both production and decay branching ratios ($\sigma \times \mathcal{B}$), accounting for all specified final states in Table~\ref{cross-sm}.
\begin{table}[htb]
\centering %
\caption{Cross sections (in pb) for SM background processes, showing production cross sections multiplied by the relevant branching ratios ($\sigma \times \mathcal{B}$) for all specified decay channels. \label{cross-sm}}
\vspace{0.8cm}
\begin{tabular}{p{2.6cm}<{\centering} p{2.6cm}<{\centering} p{2.6cm}<{\centering} p{2.4cm}<{\centering}p{2.4cm}<{\centering}}
\toprule[1.5pt]
Process&Decay Channel&$\sqrt{s}= 5.29$~TeV&6.48~TeV&9.16~TeV\\ \hline
\multirow{2}{*}{$\mu p\to\nu_{\mu} tb$}&$t\to b\ell\nu$&12.8&17.6&29.7\\
&$t\to bjj$&38.2&52.9&89.1\\ \hline
\multirow{2}{*}{$\mu p\to \nu_{\mu} Wj$}&$W\to \ell\nu$&3.7 &4.76&7.16 \\
&$W\to jj$&11.1&14.3&21.5\\ \hline
$\mu p\to \nu_{\mu} Zj$&$Z\to q\bar{q},b\bar{b}$&5.37&6.98&10.7\\ \hline
$\mu p\to \nu_{\mu}jj$ &...&378&480&701.4\\
\hline
\end{tabular}
 \end{table}

Signal and background events are generated at LO  using
MadGraph5-aMC$@$NLO with the aforementioned PDFs. Further, parton showers and hadronization are performed using Pythia 8.3 \cite{pythia8}.  Then, Delphes 3.4.2~\cite{deFavereau:2013fsa} is used for fast detector simulation, using the standard LHeC detector card.
The anti-$k_{t}$ algorithm~\cite{Cacciari:2008gp} with a jet radius of $R=0.4$ is used for small-radius~(small-$R$) jets.
For fat jet studies, the jets are reconstructed by the Cambridge-Aachen algorithm~\cite{Dokshitzer:1997in,Wobisch:1998wt}, implemented in  the FastJet  package~\cite{Cacciari:2011ma}, assuming a cone radius $R=0.8$. The $b$-tagging efficiency is set to 80\%, while the misidentification rates are 0.1\% for light-flavor jets ($u$, $d$, $s$, $g$) and 5\% for charm jets.
 Finally, the reconstructed events are analyzed using MadAnalysis 5~\cite{Conte:2012fm,Conte:2014zja}.

To identify objects, the following basic (or generation) cuts are chosen
 \be
p_{T}^{\ell/{\rm jet}}>~15~\gev,\quad
 |\eta_{\ell}|<~2.5, \quad
  |\eta_{\rm jet}|<~5, \quad
 \Delta R_{xy} > 0.4,\\
  \ee
where $p_{T}^{\ell/{\rm jet}}$ and $|\eta_{\ell/{\rm jet}}|$ are the transverse momentum and pseudorapidity of the electrons ($\ell=e$) and jets $b$ and $J$. Here, $\Delta R(x,y)=\sqrt{\Delta\Phi^{2}+\Delta\eta^{2}}$ is the separation in the pseudorapidity-azimuth plane between the pairs of objects $x$ and $y$, where $x,y=e, b, J$, wherein $b$ represents a $b$-tagged jet.

\subsubsection{Analysis of signal events in the $W_{\text{lep}}$ channel}
The following three signal benchmark points are taken with the fixed parameter $g^{*}=0.1$:
\begin{enumerate}[(i)]
\item
$T_{1500}$:~ $m_{T}=1500$ GeV;
\item
 $T_{2000}$:~$ m_{T}=2000$ GeV;
 \item $T_{2500}$: $m_{T}=2500$ GeV.
\end{enumerate}
%
\begin{figure*}[htb]
\begin{center}
\centerline{\hspace{2.0cm}\epsfxsize=8cm\epsffile{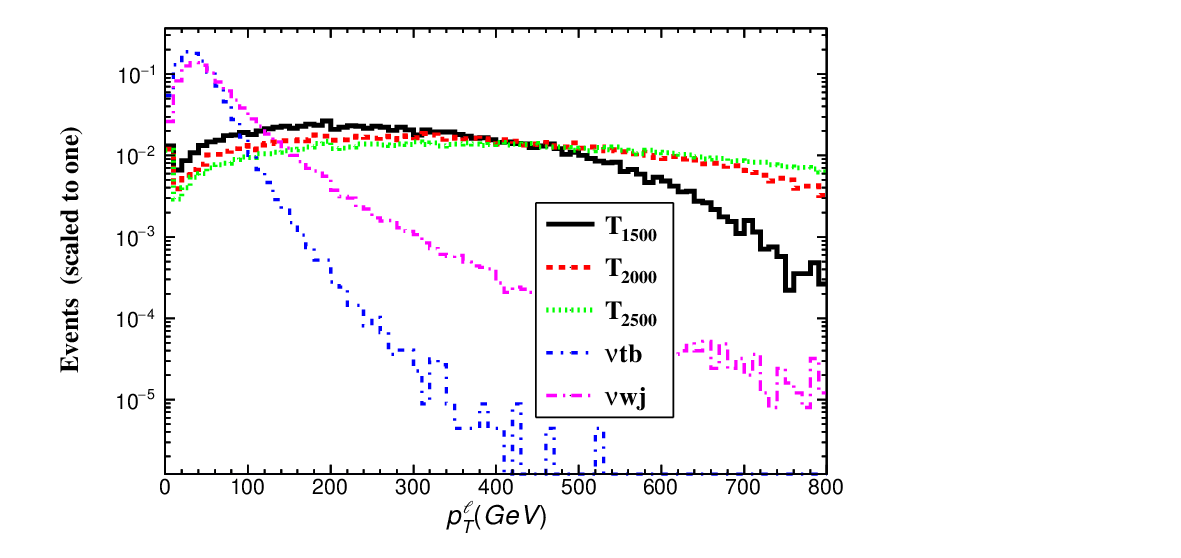}
\hspace{-2.0cm}\epsfxsize=8cm\epsffile{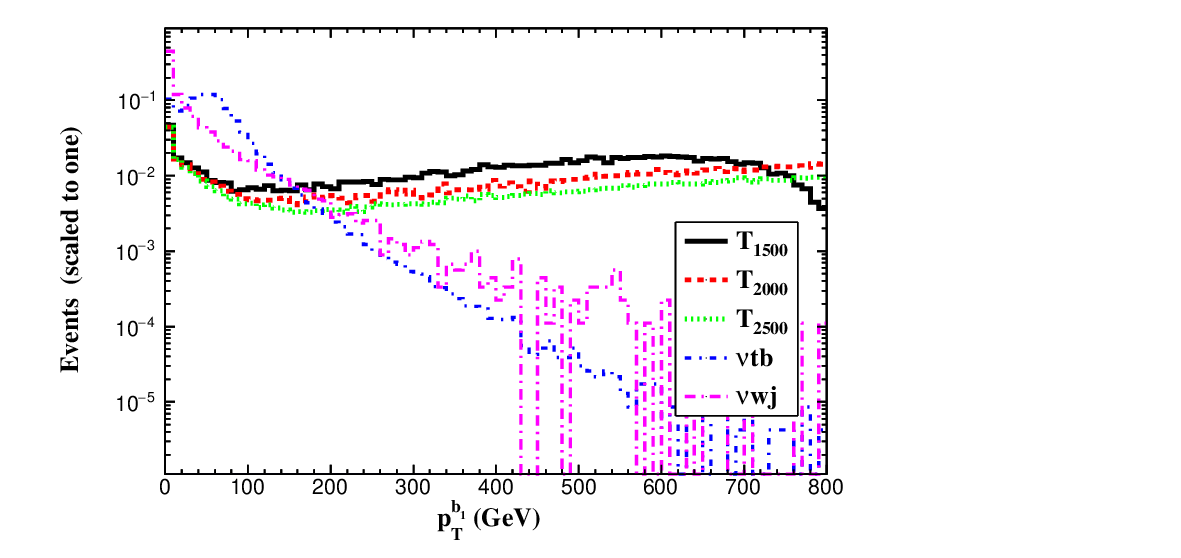}}
\centerline{\hspace{2.0cm}\epsfxsize=8cm\epsffile{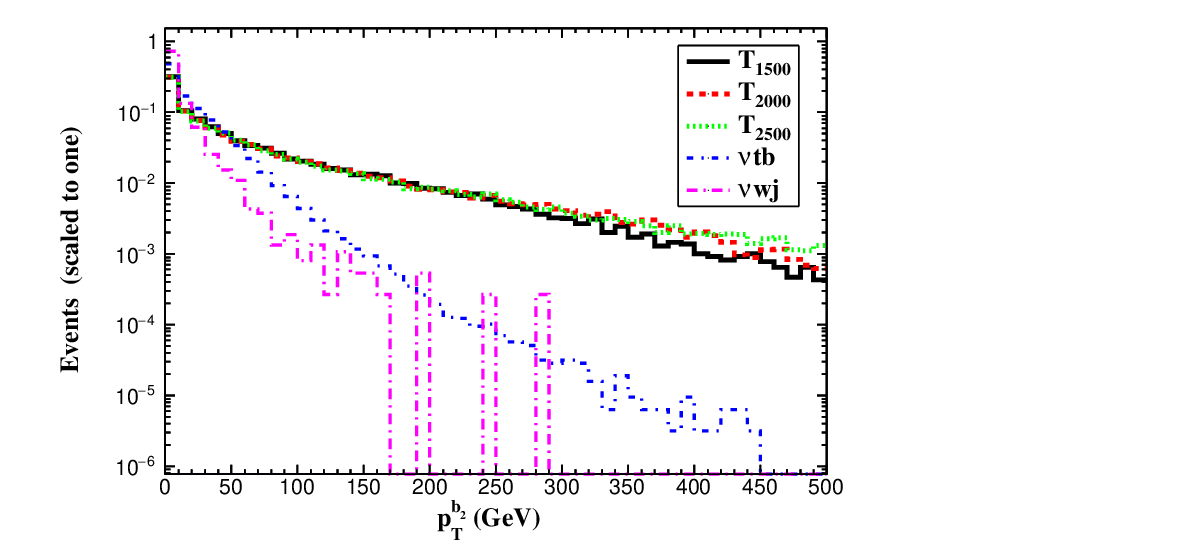}
\hspace{-2.0cm}\epsfxsize=8cm\epsffile{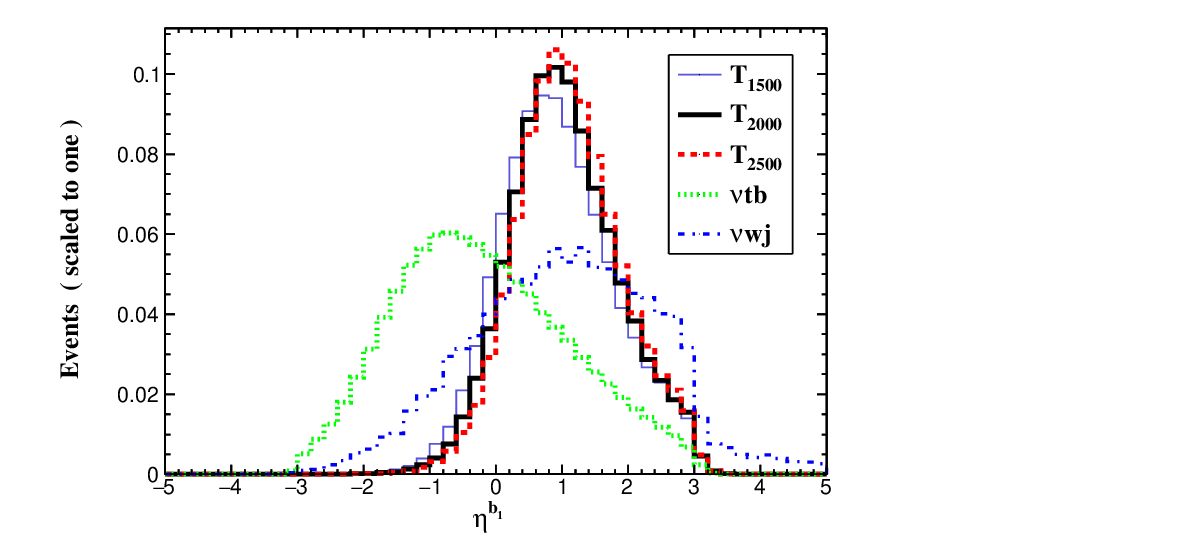}}
\centerline{\hspace{2.0cm}\epsfxsize=8cm\epsffile{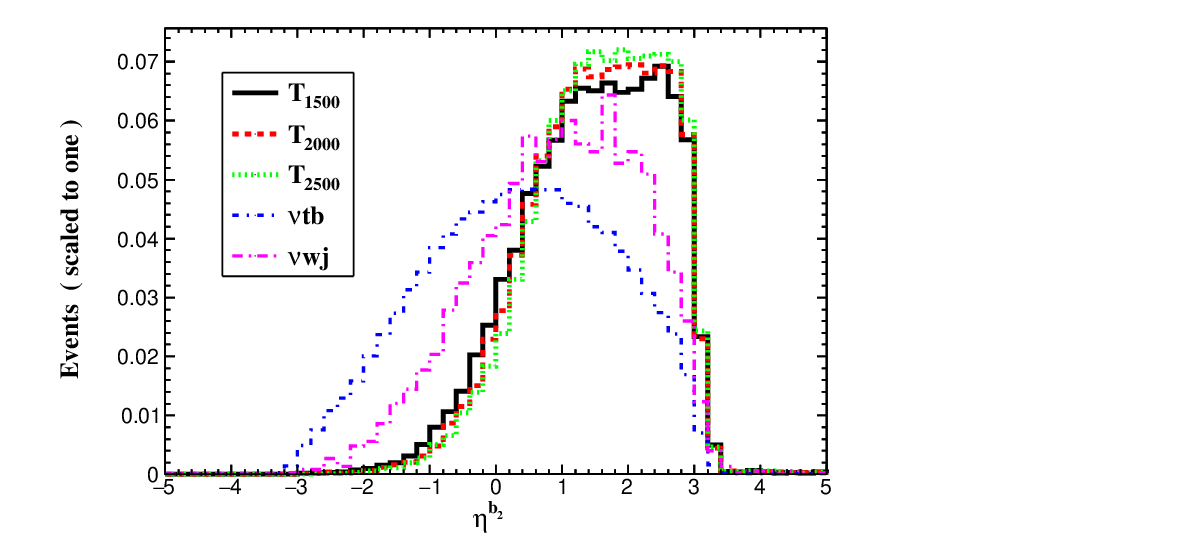}
\hspace{-2.0cm}\epsfxsize=8cm\epsffile{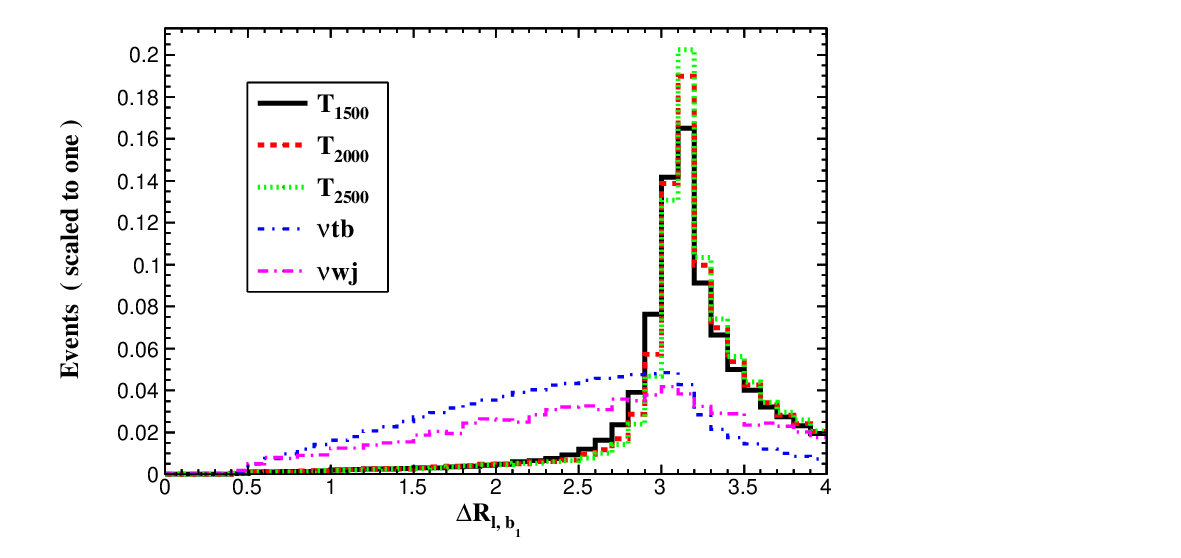}}
\centerline{\hspace{2.0cm}\epsfxsize=8cm\epsffile{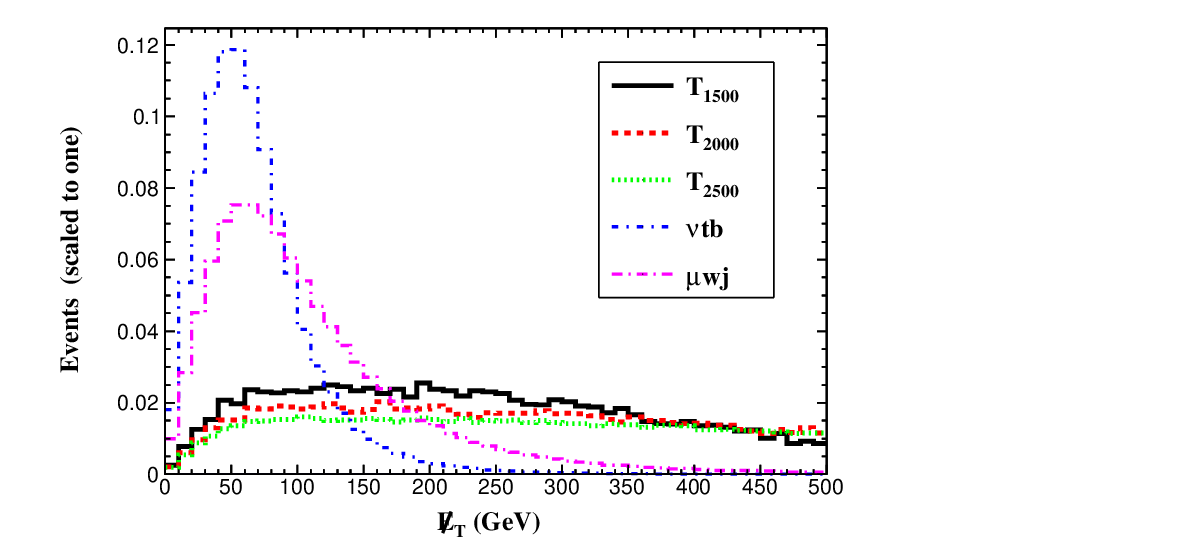}
\hspace{-2.0cm}\epsfxsize=8cm\epsffile{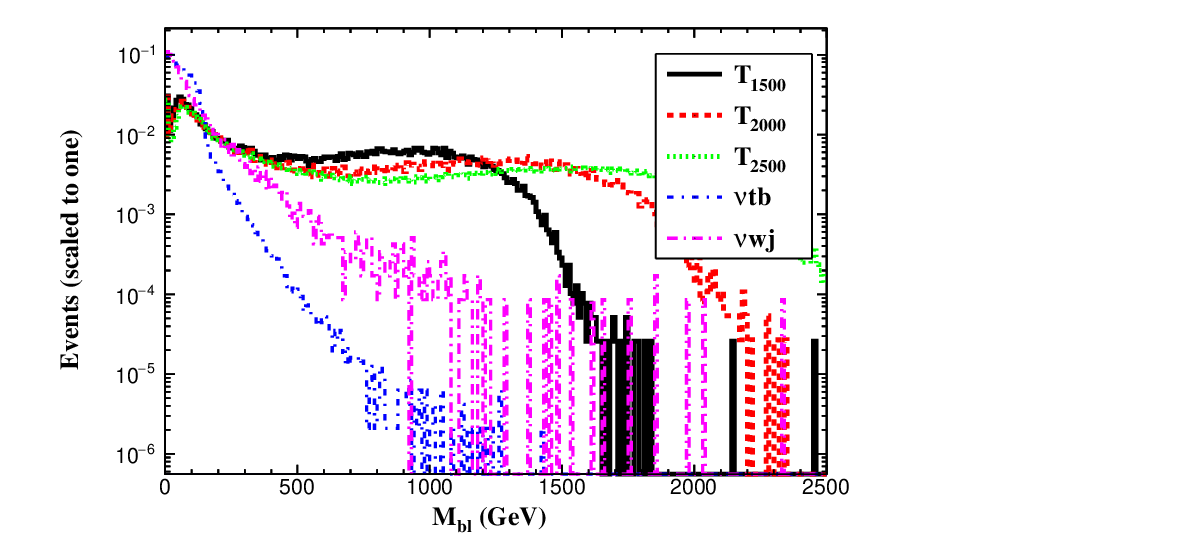}}
\caption{Normalized distributions of $p_T^\ell$, $p_T^{b_{1,2}}$, $\eta^{b_{1,2}}$, $\Delta R_{e,b_{1}}$, $\slashed{E}_T$, $M_{be}$ for the three signals [with $m_T = 1500$ GeV (solid), 2000 GeV (dashed), and 2500 GeV (dotted)] and SM backgrounds at $\sqrt{s}= 5.29$~TeV in the $W_{\text{lep}}$ channel. }
\label{fig4}
\end{center}
\end{figure*}

In Fig.~\ref{fig4}, we plot some differential distributions  for the three signal benchmark points~($T_{1500}$, $T_{2000}$ and $T_{2500}$) and SM backgrounds at $\sqrt{s}= 5.29$~TeV in the $W_{\text{lep}}$ channel, including the transverse momentum distributions of the lepton~($p_{T}^{\ell}$), the transverse momentum~($p_{T}^{b_{1,2}}$) and pseudorapidity~($\eta^{b_{1,2}}$) distributions for both the leading and subleading $b$-tagged jets, the pseudorapidity-azimuth separation between the leading $b$-tagged jet and the lepton of $\Delta R_{\ell,b_{1}}$, the missing transverse energy $\slashed E_{T}$ and the invariant mass distribution $M_{b e}$. Because of the larger mass of VLQ-$T$, the decay products are highly energetic, and thus the $p_{T}^{e}$ and $p_{T}^{b}$ peaks of the signals are larger than those of the SM backgrounds.
According to the behaviors of
these distributions, we impose
the following cuts to distinguish the signal from the SM backgrounds.
 \begin{itemize}
\item Cut 1: Exactly one isolated electron with  $p_{T}^e> 150 \rm ~GeV$ (i.e., events with
final state muons  or taus are vetoed).
\item Cut 2: Events must contain at least one b-tagged jet ($N_b \geq 1$). The selected $b$-jet (defined as the highest-$p_{T}$ $b$-tagged jet when multiple candidates exist) must satisfy $p_{T}^{b} > 200~\mathrm{GeV}$ and maintain a pseudorapidity-azimuth separation from the lepton of $\Delta R_{e,b} > 2.7$.
\item  Cut 3: The transverse missing energy is required to be  $\slashed E_{T}> 300 \rm ~GeV$.
\item Cut 4: The $be$ invariant mass is required to be $M_{b e}> 1000 \rm ~GeV$.
\end{itemize}

Since the
behaviors of the relevant kinematic distributions at $\sqrt{s}= 6.48$ and 9.16~TeV are similar to the case of $\sqrt{s}= 5.29$~TeV, we do not display these here. Based on the behaviors of these distributions, we impose the same cuts to enhance the sensitivity also at other two c.m. energies.

\begin{table}[htb]
\centering %
\caption{Cut flow of the cross sections (in fb) for the signals and SM backgrounds at $\sqrt{s}= 5.29$~TeV in the $W_{\text{lep}}$ channel. Here, we set a benchmark value of $g^{*}=0.1$. \label{cutflow-bl1}}
\vspace{0.8cm}
\begin{tabular}{p{2.0cm}<{\centering} p{1.3cm}<{\centering} p{1.3cm}<{\centering} p{1.3cm}<{\centering}p{0.1cm}<{\centering} p{1.6cm}<{\centering}  p{1.6cm}<{\centering}  }
\toprule[1.5pt]
 \multirow{2}{*}{Cuts}& \multicolumn{3}{c}{Signals}&&\multicolumn{2}{c}{Backgrounds}  \\ \cline{2-4} \cline{6-7}
&$T_{1500}$&$T_{2000}$ &$T_{2500}$&& $\nu tb$& $\nu Wj$  \\    \cline{1-7} \midrule[1pt]
Basic&0.35&0.074&0.014&&7296&2331\\
Cut 1&0.12&0.028&0.0054&&26&102\\
Cut 2&0.084&0.02&0.004&&5.1&0.42\\
Cut 3 &0.06&0.015&0.0032&&0.84&0.13\\
Cut 4 &0.024&0.012&0.003&&0.077&0.037\\ \hline
Total eff.&0.06&0.14&0.18&&6.0E-6&1.0E-5\\
\hline
\end{tabular}
 \end{table}

\begin{table}[htb]
\centering %
\caption{Cut flow of the cross sections (in fb) for the signals and SM backgrounds at $\sqrt{s}= 6.48$~TeV in the $W_{\text{lep}}$ channel. Here, we set a benchmark value of $g^{*}=0.1$. \label{cutflow-bl2}}
\vspace{0.8cm}
\begin{tabular}{p{1.4cm}<{\centering} p{1.3cm}<{\centering} p{1.3cm}<{\centering} p{1.3cm}<{\centering}p{0.1cm}<{\centering} p{1.6cm}<{\centering}  p{1.6cm}<{\centering} p{1.6cm}<{\centering}}
\toprule[1.5pt]
 \multirow{2}{*}{Cuts}& \multicolumn{3}{c}{Signals}&&\multicolumn{2}{c}{Backgrounds}  \\ \cline{2-4} \cline{6-7}
&$T_{1500}$&$T_{2000}$ &$T_{2500}$&& $\nu tb$ & $\nu Wj$ \\    \cline{1-7} \midrule[1pt]
Basic&0.84&0.24&0.065&&9328&2761\\
Cut 1&0.29&0.09&0.026&&41&143\\
Cut 2&0.19&0.065&0.018&&8.8&0.51\\
Cut 3 &0.14&0.05&0.015&&1.62&0.22\\
Cut 4 &0.06&0.04&0.013&&0.28&0.076\\ \hline
Total eff.&0.06&0.15&0.19&&1.6E-5&1.6E-5\\
\hline
\end{tabular}
 \end{table}

 \begin{table}[htb]
\centering %
\caption{Cut flow of the cross sections (in fb) for the signals and SM backgrounds at $\sqrt{s}= 9.16$~TeV in the $W_{\text{lep}}$ channel. Here, we set a benchmark value of $g^{*}=0.1$.  \label{cutflow-bl3}}
\vspace{0.8cm}
\begin{tabular}{p{1.4cm}<{\centering} p{1.3cm}<{\centering} p{1.3cm}<{\centering} p{1.3cm}<{\centering}p{0.1cm}<{\centering} p{1.6cm}<{\centering}  p{1.6cm}<{\centering} p{1.6cm}<{\centering} }
\toprule[1.5pt]
 \multirow{2}{*}{Cuts}& \multicolumn{3}{c}{Signals}&&\multicolumn{2}{c}{Backgrounds}  \\ \cline{2-4} \cline{6-7}
&$T_{1500}$&$T_{2000}$ &$T_{2500}$&& $\nu tb$& $\nu Wj$  \\    \cline{1-7} \midrule[1pt]
Basic&2.72&1.03&0.41&&13055&3437\\
Cut 1&0.96&0.4&0.16&&82&208\\
Cut 2&0.67&0.28&0.12&&18&0.87\\
Cut 3 &0.48&0.22&0.11&&4.3&0.57\\
Cut 4 &0.21&0.18&0.09&&0.27&0.37\\ \hline
Total eff.&0.063&0.15&0.19&&9.0E-6&5.2E-5\\
\hline
\end{tabular}
 \end{table}

The cross sections of the three typical signals and the relevant
SM backgrounds are presented in Tables~\ref{cutflow-bl1}-\ref{cutflow-bl3} after imposing
the aforementioned cuts.  The cross sections for signal and SM backgrounds are calculated by
$\sigma_{\text{after cut}} = \sigma_0 \times \epsilon_{\text{cut}}$,
where $\sigma_0$ denotes the initial cross section and $\epsilon_{\text{cut}}$ represents the cumulative efficiency up to that cut. The efficiency for an individual cut can then be obtained from the ratio of consecutive cross section values in the table. The values in the last column of the cut-flow tables represent the total cumulative efficiencies for both signal and background processes. The appendix include the details about the events and efficiencies before and after each cut at $\sqrt{s}= 5.29$~TeV in the $W_{\text{lep}}$ channel.
Notably,  all background processes are very significantly suppressed  at the end of the cut flow, and the cross section of the total SM  background  is about 0.11~fb at at $\sqrt{s}= 5.29$~TeV, 0.36 fb at $\sqrt{s}= 6.48$~TeV, and 0.64 fb at $\sqrt{s}= 9.16$~TeV, respectively.

\subsubsection{Analysis of signal events in the $W_{\text{had}}$ channel}
\begin{figure*}[htb]
\begin{center}
\centerline{\hspace{2.0cm}\epsfxsize=9cm\epsffile{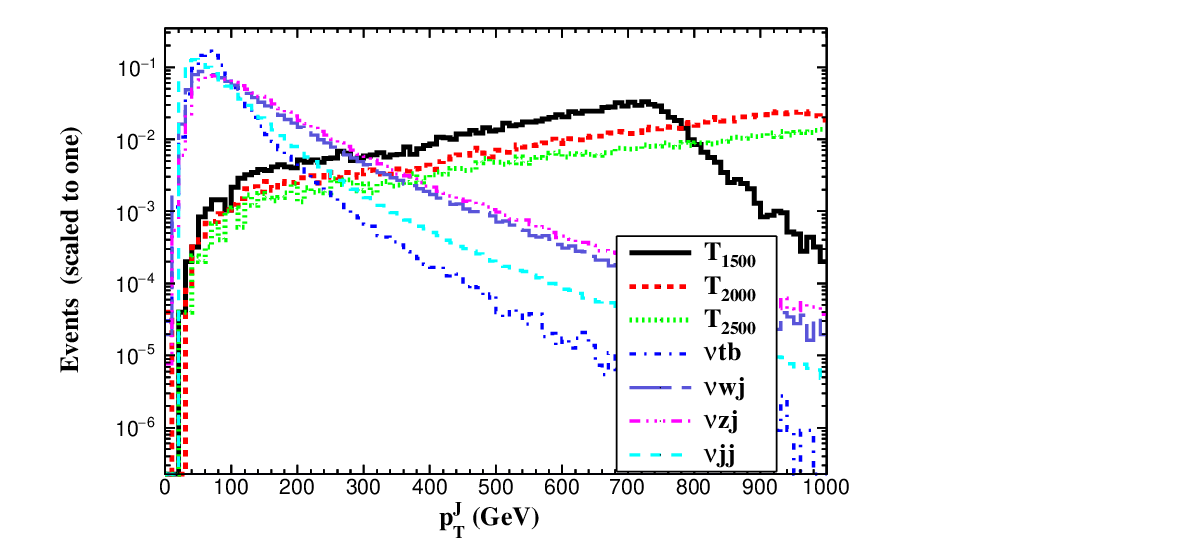}
\hspace{-2.0cm}\epsfxsize=9cm\epsffile{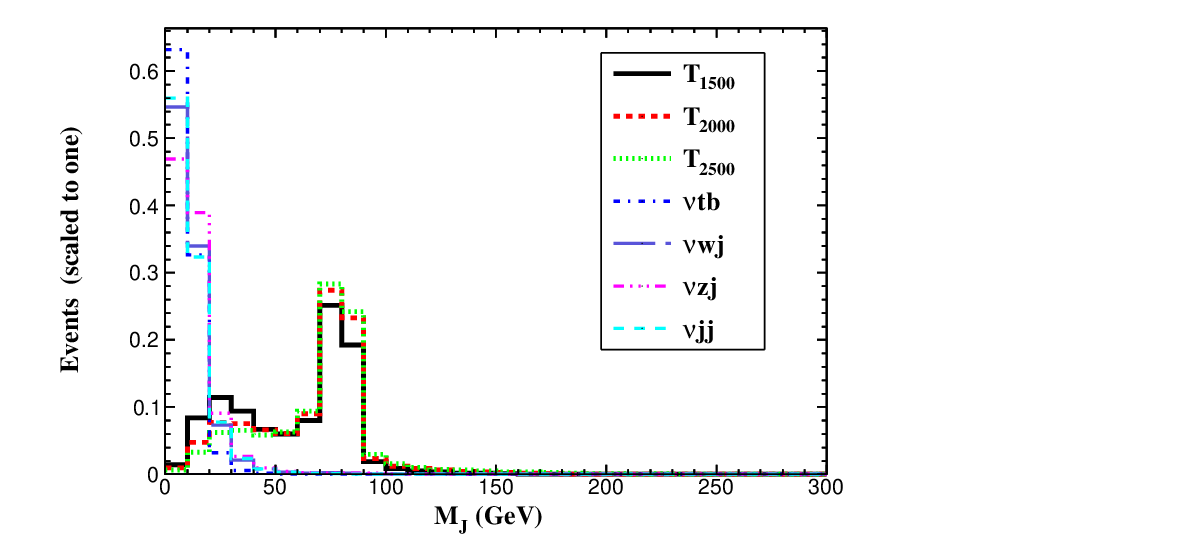}}
\centerline{\hspace{2.0cm}\epsfxsize=9cm\epsffile{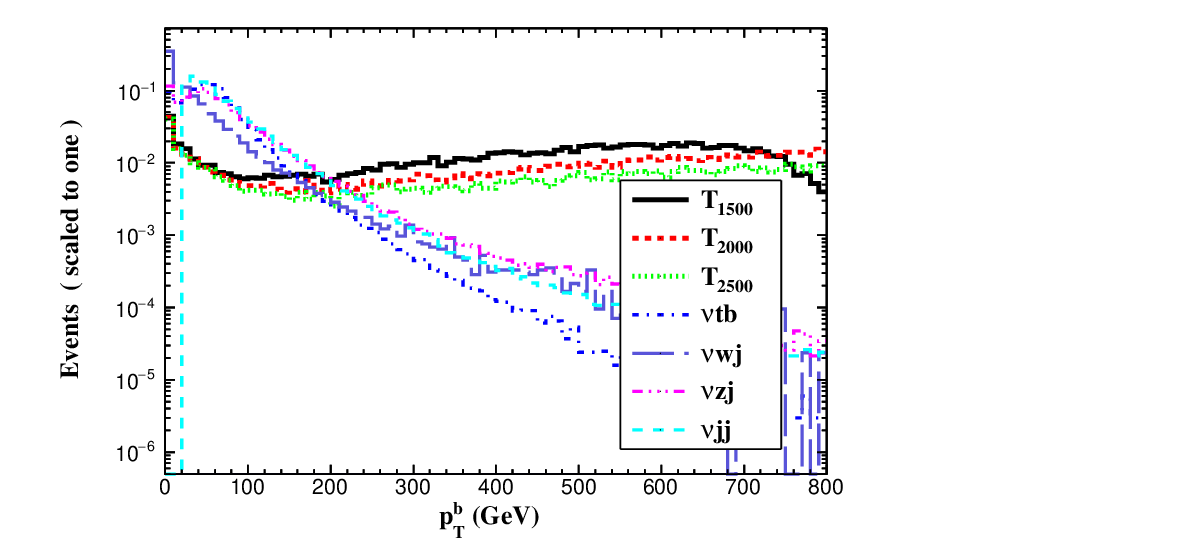}
\hspace{-2.0cm}\epsfxsize=9cm\epsffile{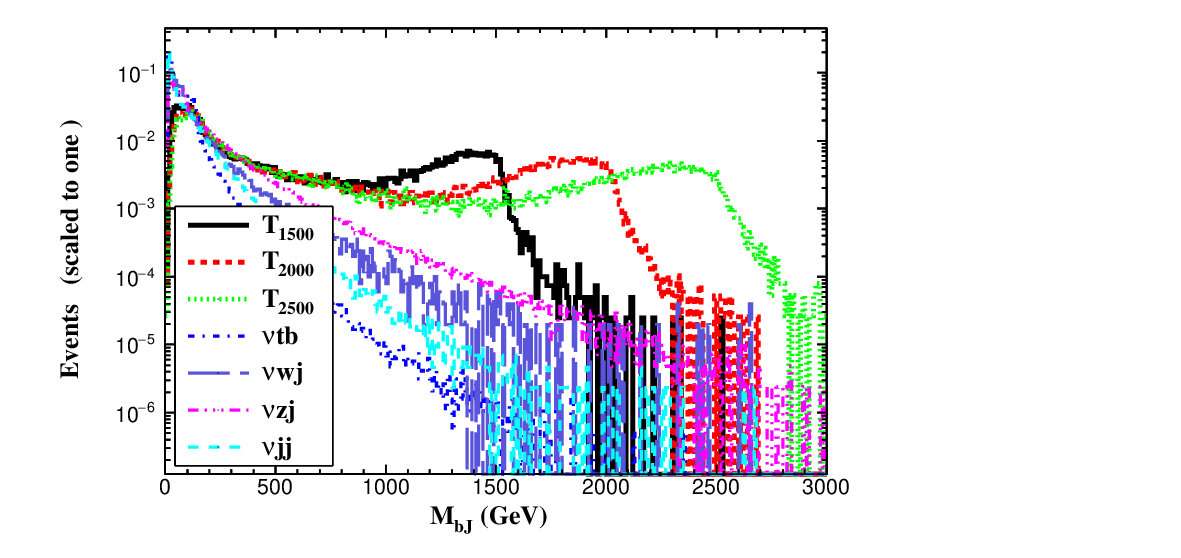}}
\caption{Normalized distributions of $p_T^J$, $M_{J}$, $p_T^{b}$, $M_{bJ}$ for the signals (with $m_{T}=1500, 2000$, and 2500 GeV) and SM backgrounds  at $\sqrt{s}= 5.29$~TeV in the $W_{\text{had}}$ channel.  }
\label{fig5}
\end{center}
\end{figure*}

In Fig.~\ref{fig5}, we show the normalized distributions of  the transverse momentum  of the fat jet~($p_{T}^{J}$) and leading $b$-tagged jet~($p_{T}^{b}$) as well as those of  the fat jet mass~($M_{J}$) and of the invariant mass of the $b$-tagged  and fat jet system~($M_{bJ}$). (Again, these are presented for $\sqrt{s}= 5.29$~TeV in the $W_{\text{had}}$ channel, but they are very similar for $\sqrt{s}= 6.48$~TeV and 9.16~TeV.)
According to the behaviors of
these distributions, we impose
the following cuts to extract the signal from the SM backgrounds.
 \begin{itemize}
\item Cut 1: The transverse momentum for the fat jet is required to be $p_{T}^{J}> 200 \rm ~GeV$.
\item Cut 2: The fat jet mass is such that $|M_{J}-m_{W}| <15 \rm ~GeV$.
\item Cut 3: Events are required to have at least one $b$-tagged jet ($N_{b} \geq 1$). The leading $b$-tagged jet must satisfy $p_{T}^{b} > 200~\mathrm{GeV}$, and its separation from the lepton in pseudorapidity-azimuth space must satisfy $\Delta R_{e,b} > 2.7$.
\item Cut 4: The aforementioned  invariant mass is required to be: $M_{bJ}> 1300 \rm ~GeV$ for $m_{T}<1800~\rm GeV$~(Cut 4a), and $M_{bJ}> 1500 \rm ~GeV$ for $m_{T}\geq1800~\rm GeV$~(Cut 4b).
\end{itemize}

\begin{table}[htb]
\centering %
\caption{Cut flow of the cross sections (in fb) for the signals and SM backgrounds at $\sqrt{s}= 5.29$~TeV in the $W_{\text{had}}$ channel. Here, we set a benchmark value of $g^{*}=0.1$. \label{cutflow-fat1}}
\vspace{0.8cm}
\begin{tabular}{p{2.4cm}<{\centering} p{1.3cm}<{\centering} p{1.3cm}<{\centering} p{1.3cm}<{\centering}p{0.1cm}<{\centering} p{1.6cm}<{\centering}  p{1.6cm}<{\centering} p{1.6cm}<{\centering} p{1.6cm}<{\centering} }
\toprule[1.5pt]
 \multirow{2}{*}{Cuts}& \multicolumn{3}{c}{Signals}&&\multicolumn{3}{c}{Backgrounds}  \\ \cline{2-4} \cline{6-9}
&$T_{1500}$&$T_{2000}$ &$T_{2500}$&& $\nu tb$& $\nu Wj$& $\nu Zj$&  $\nu jj$  \\    \cline{1-9} \midrule[1pt]
Basic&1.21&0.25&0.05&&38160&11100&5365&378000\\
Cut 1&1.16&0.25&0.046&&954&1576&268&17010\\
Cut 2&0.61&0.15&0.03&&23&68&11&2438\\
Cut 3 &0.39&0.094&0.018&&2.67&0.33&0.53&2.04\\
Cut 4a &0.242&$\backslash$&$\backslash$&&0.106&0.054&0.183&0.216\\
Cut 4b &$\backslash$&0.078&0.017&&0.021&0.053&0.125&0.087\\ \hline
Total eff.$_{\text{Cut-4a}}$  &20\%&$\backslash$&$\backslash$&&2.8E-6&4.9E-6&3.4E-5&5.7E-7\\
Total eff.$_{\text{Cut-4b}}$ &$\backslash$&31\%&35\%&&5.6E-7&4.8E-6&2.3E-5&2.3E-7\\
\hline
\end{tabular}
 \end{table}

\begin{table}[htb]
\centering %
\caption{Cut flow of the cross sections (in fb) for the signals and SM backgrounds at $\sqrt{s}= 6.48$~TeV in the $W_{\text{had}}$ channel. Here, we set a benchmark value of $g^{*}=0.1$. \label{cutflow-fat2}}
\vspace{0.8cm}
\begin{tabular}{p{2.4cm}<{\centering} p{1.3cm}<{\centering} p{1.3cm}<{\centering} p{1.3cm}<{\centering}p{0.1cm}<{\centering} p{1.6cm}<{\centering}  p{1.6cm}<{\centering} p{1.6cm}<{\centering} p{1.6cm}<{\centering} }
\toprule[1.5pt]
 \multirow{2}{*}{Cuts}& \multicolumn{3}{c}{Signals}&&\multicolumn{3}{c}{Backgrounds}  \\ \cline{2-4} \cline{6-9}
&$T_{1500}$&$T_{2000}$ &$T_{2500}$&& $\nu tb$& $\nu Wj$& $\nu Zj$ & $\nu jj$  \\    \cline{1-9} \midrule[1pt]
Basic&2.88&0.81&0.22&&52920&14280&6980&480000\\
Cut 1&2.76&0.78&0.22&&1534&2142&138&23520\\
Cut 2&1.44&0.46&0.13&&45&114&58&377\\
Cut 3 &0.86&0.29&0.084&&4.76&0.61&0.77&0.53\\
Cut 4a &0.55&$\backslash$&$\backslash$&&0.40&0.36&0.31&0.17\\
Cut 4b &$\backslash$&0.24&0.08&&0.167&0.248&0.24&0.082\\ \hline
Total eff.$_{\text{Cut-4a}}$  &19\%&$\backslash$&$\backslash$&&7.6E-6&2.5E-5&4.5E-5&3.5E-7\\
Total eff.$_{\text{Cut-4b}}$  &$\backslash$&30\%&36\%&&3.2E-6&1.7E-5&3.5E-5&1.7E-7\\
\hline
\end{tabular}
 \end{table}

\begin{table}[htb]
\centering %
\caption{Cut flow of the cross sections (in fb) for the signals and SM backgrounds at $\sqrt{s}= 9.16$~TeV in the $W_{\text{had}}$ channel. Here, we set a benchmark value of $g^{*}=0.1$. \label{cutflow-fat3}}
\vspace{0.8cm}
\begin{tabular}{p{2.4cm}<{\centering} p{1.3cm}<{\centering} p{1.3cm}<{\centering} p{1.3cm}<{\centering}p{0.1cm}<{\centering} p{1.6cm}<{\centering}  p{1.6cm}<{\centering}  p{1.6cm}<{\centering}p{1.6cm}<{\centering}}
\toprule[1.5pt]
 \multirow{2}{*}{Cuts}& \multicolumn{3}{c}{Signals}&&\multicolumn{3}{c}{Backgrounds}  \\ \cline{2-4} \cline{6-8}
&$T_{2000}$&$T_{3000}$ &$T_{4000}$&& $\nu tb$& $\nu Wj$& $\nu Zj$  & $\nu jj$  \\    \cline{1-8} \midrule[1pt]
Basic&3.62&0.58&0.0884&&89000&21480&10700&701400\\
Cut 1&3.54&0.58&0.088&&3204&3651&2354&38577\\
Cut 2&2.32&0.4&0.062&&196&301&107&982\\
Cut 3 &1.22&0.232&0.036&&12.5&1.85&1.07&2.1\\
Cut 4 &1.09&0.226&0.035&&0.56&0.47&0.39&0.22\\\hline
Total eff. &28\%&36.5\%&38.2\%&&6.3E-6&2.2E-5&3.7E-5&3.2E-7\\
\hline
\end{tabular}
 \end{table}

In Tables~\ref{cutflow-fat1}-\ref{cutflow-fat3}, we
 present the cross sections for the signals and relevant
SM backgrounds after imposing
the above cuts.  Here, one can see that all the SM backgrounds are suppressed very efficiently, while the signals still have  relatively good efficiency at the end of the cut flow. The cross section of the  total SM background in the $W_{\text{had}}$ channel is about 0.28 fb at $\sqrt{s}= 5.29$~TeV, 0.52 fb at $\sqrt{s}= 6.48$~TeV, and 1.65~fb at $\sqrt{s}= 9.16$~TeV, respectively.
\subsubsection{Discovery and exclusion significance }
In order to analyze the sensitivity, we  estimate the expected discovery~($\mathcal{Z}_\text{dis}$) and exclusion~($\mathcal{Z}_\text{exc}$) limits by using the following formulas~\cite{Cowan:2010js}:
\be
\mathcal{Z}_\text{dis} &=
  \sqrt{2\left[(s+b)\ln\left(\frac{(s+b)(1+\delta_{sys}^2 b)}{b+\delta_{sys}^2 b(s+b)}\right) -
  \frac{1}{\delta_{sys}^2 }\ln\left(1+\frac{\delta_{sys}^2 s}{1+\delta_{sys}^2 b}\right)\right]}, \\
   \mathcal{Z}_\text{exc} &=\sqrt{2\left[s-b\ln\left(\frac{b+s+x}{2b}\right)
  - \frac{1}{\delta_{sys}^2 }\ln\left(\frac{b-s+x}{2b}\right)\right] -
  \left(b+s-x\right)\left(1+\frac{1}{\delta_{sys}^2 b}\right)},
 \ee
with
 \be
 x=\sqrt{(s+b)^2- 4 \delta_{sys}^2 s b^2/(1+\delta_{sys}^2 b)}.
 \ee
Here, $s$ and $b$ are the numbers of signal and background events, respectively, which can be obtained by multiplying the total signal and SM background cross
sections, respectively,  by the integrated luminosity while  $\delta$ is the percentage systematic error on the SM  background estimate.
In the limit case of $\delta_{sys} \to 0$,  these expressions  can be simplified to
\be
\mathcal{Z}_\text{dis} &= \sqrt{2[(s+b)\ln(1+s/b)-s]},\\
 \mathcal{Z}_\text{exc} &= \sqrt{2[s-b\ln(1+s/b)]}.
 \label{eq}
 \ee
The integrated luminosity
is set at 100~fb$^{-1}$ for all three center-of-mass energies~\cite{Kaya:2019ecf}.

\begin{figure}[htb]
\begin{center}
\vspace{1.5cm}
\centerline{\epsfxsize=6.5cm \epsffile{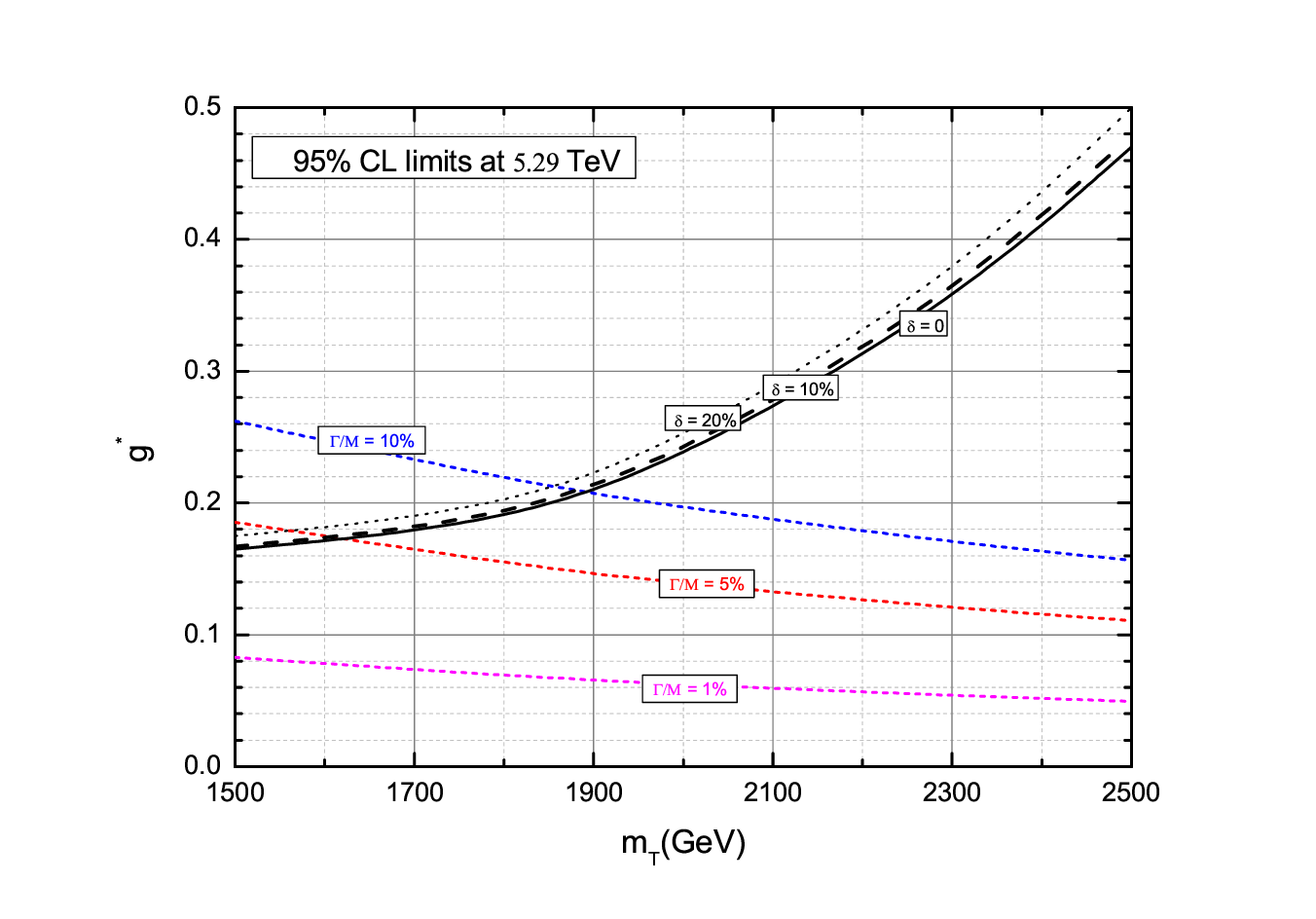}\epsfxsize=6.5cm \epsffile{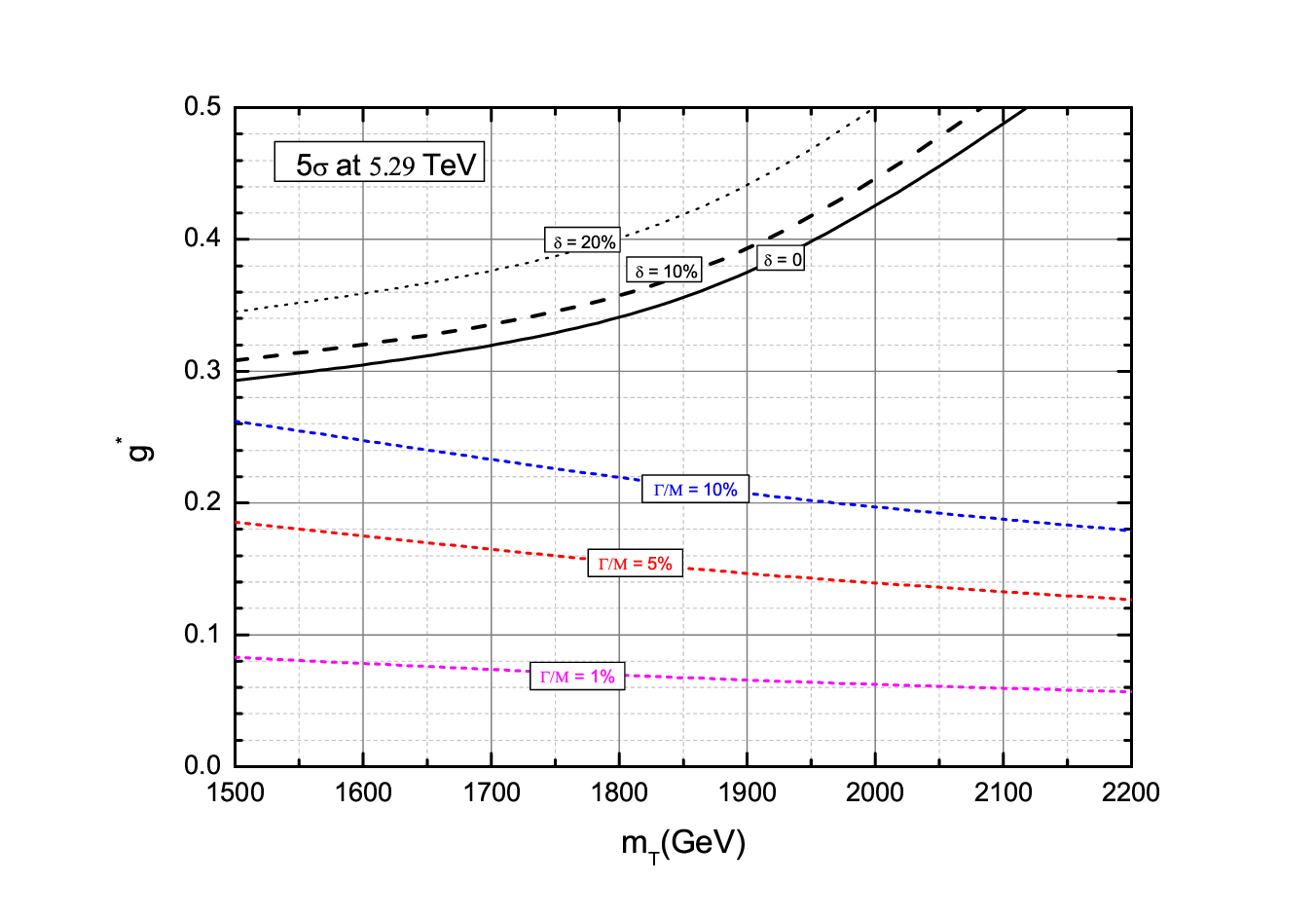}}
\centerline{\epsfxsize=6.5cm \epsffile{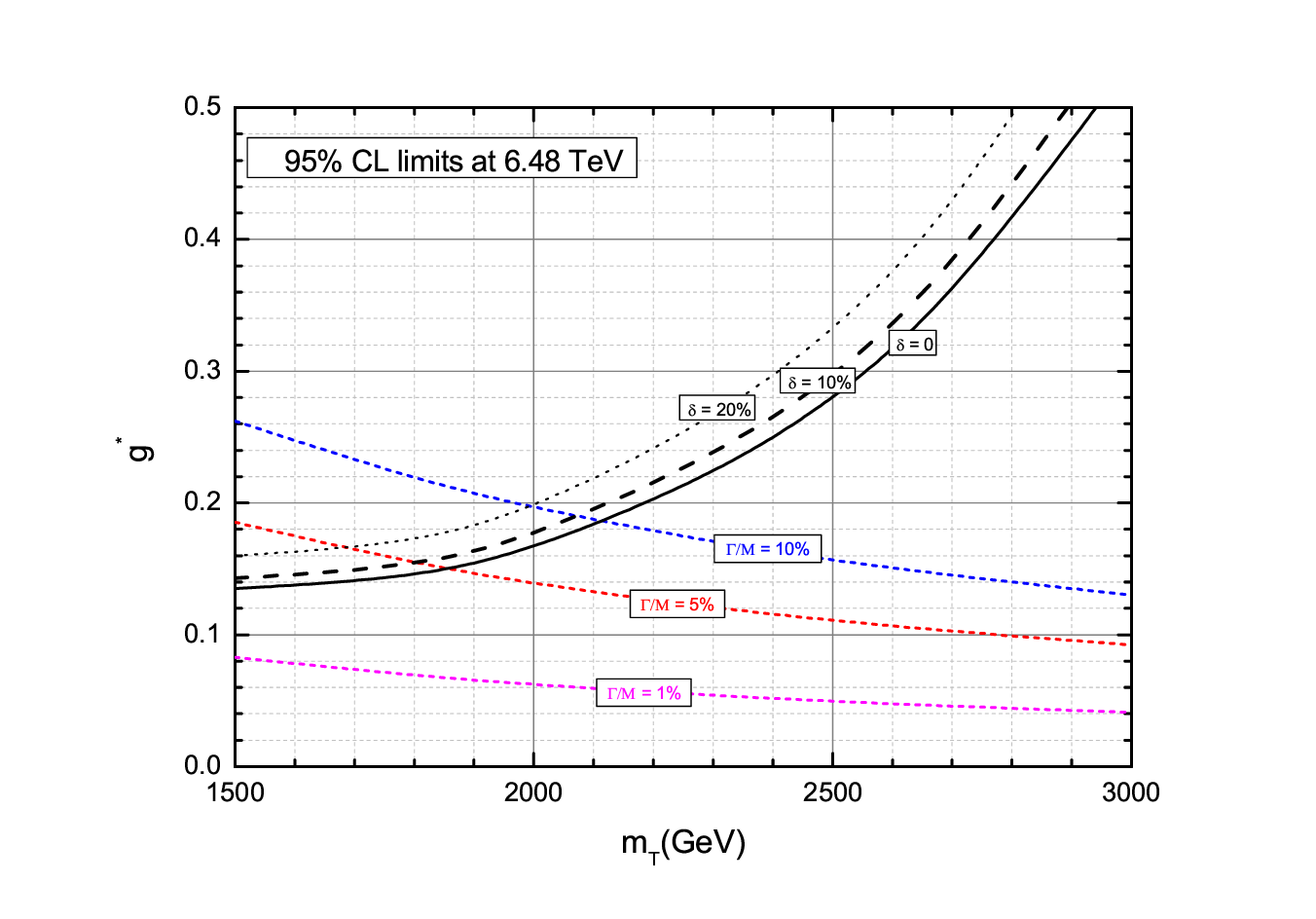}\epsfxsize=6.5cm \epsffile{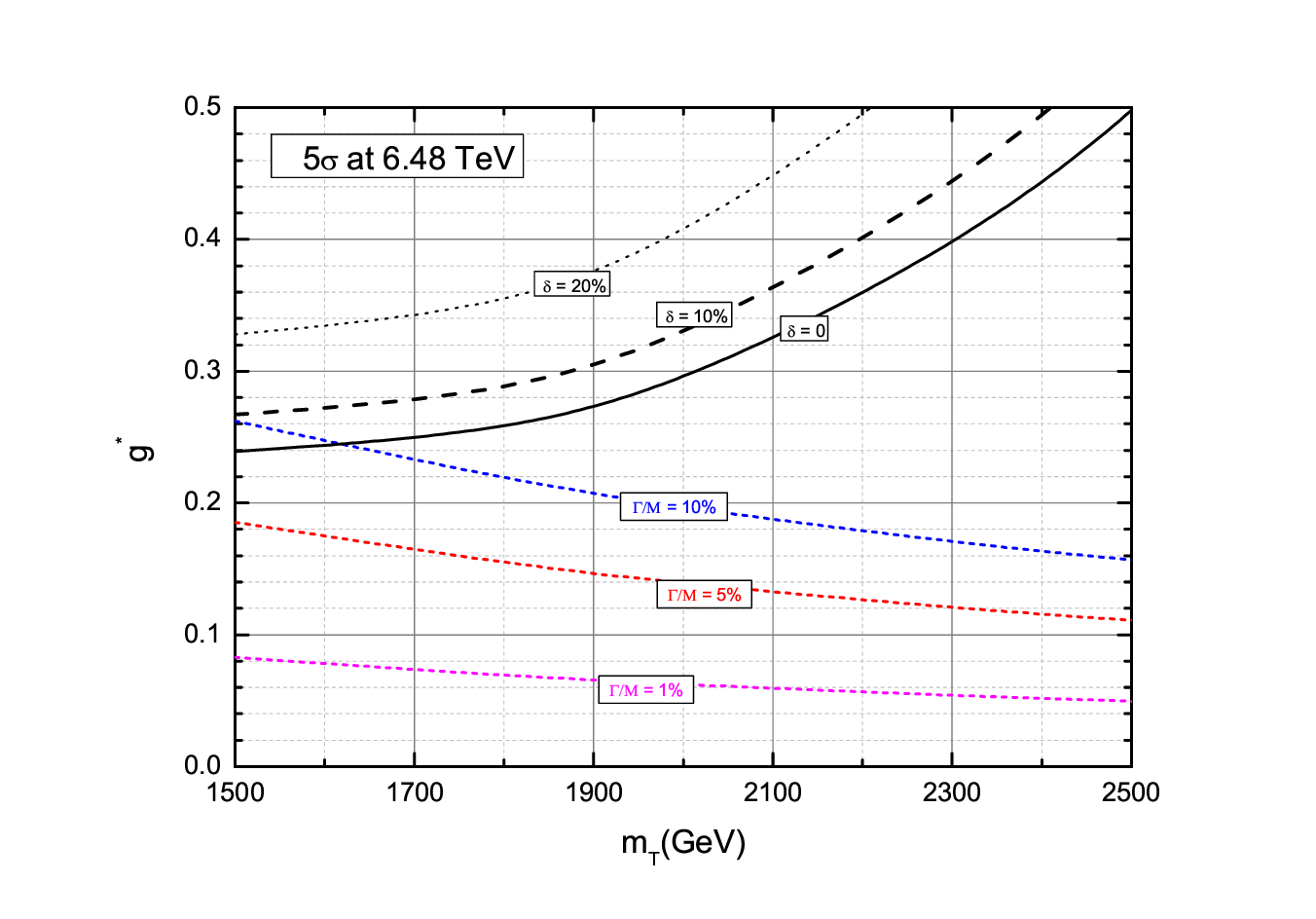}}
\centerline{\epsfxsize=6.5cm \epsffile{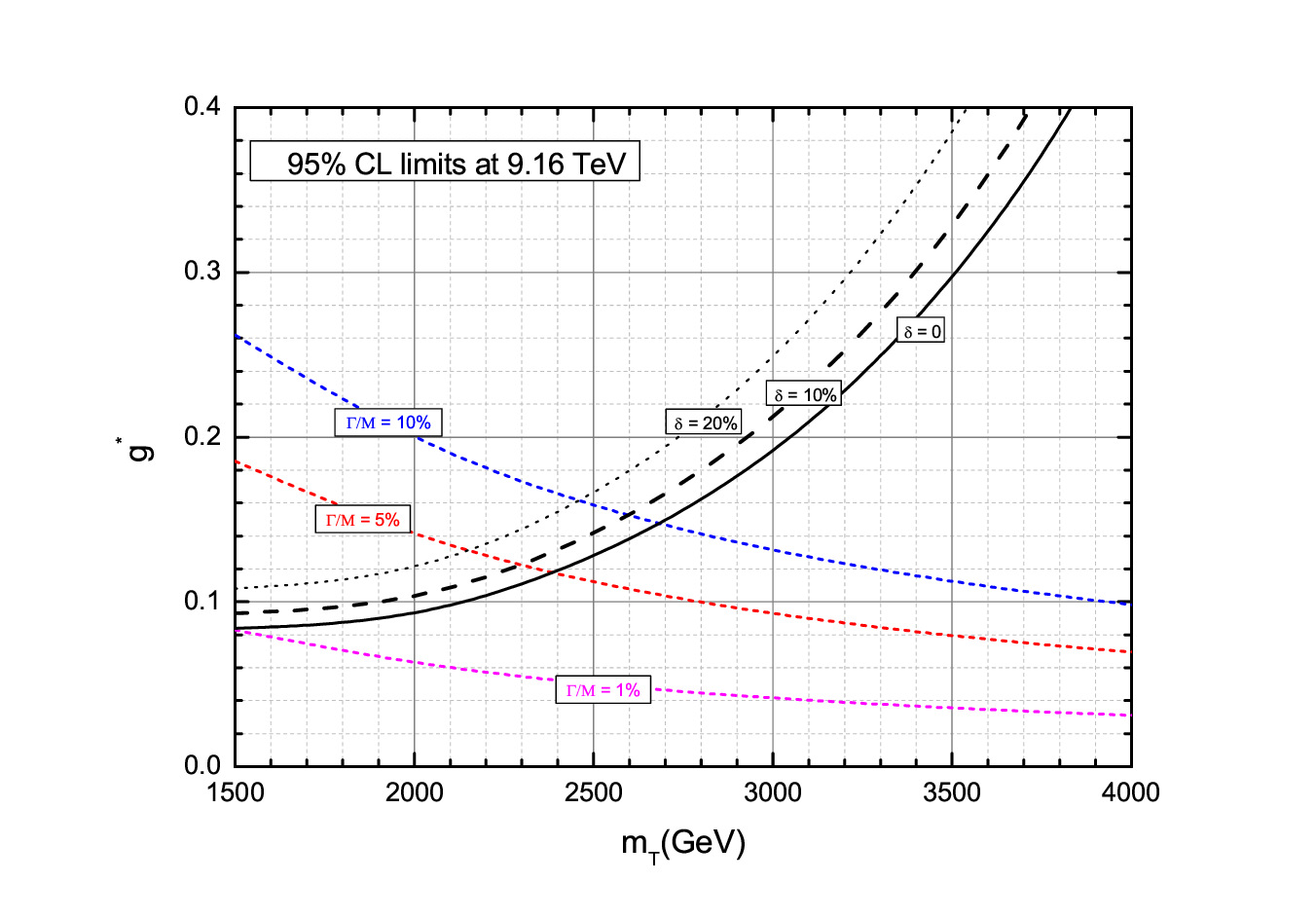}\epsfxsize=6.5cm \epsffile{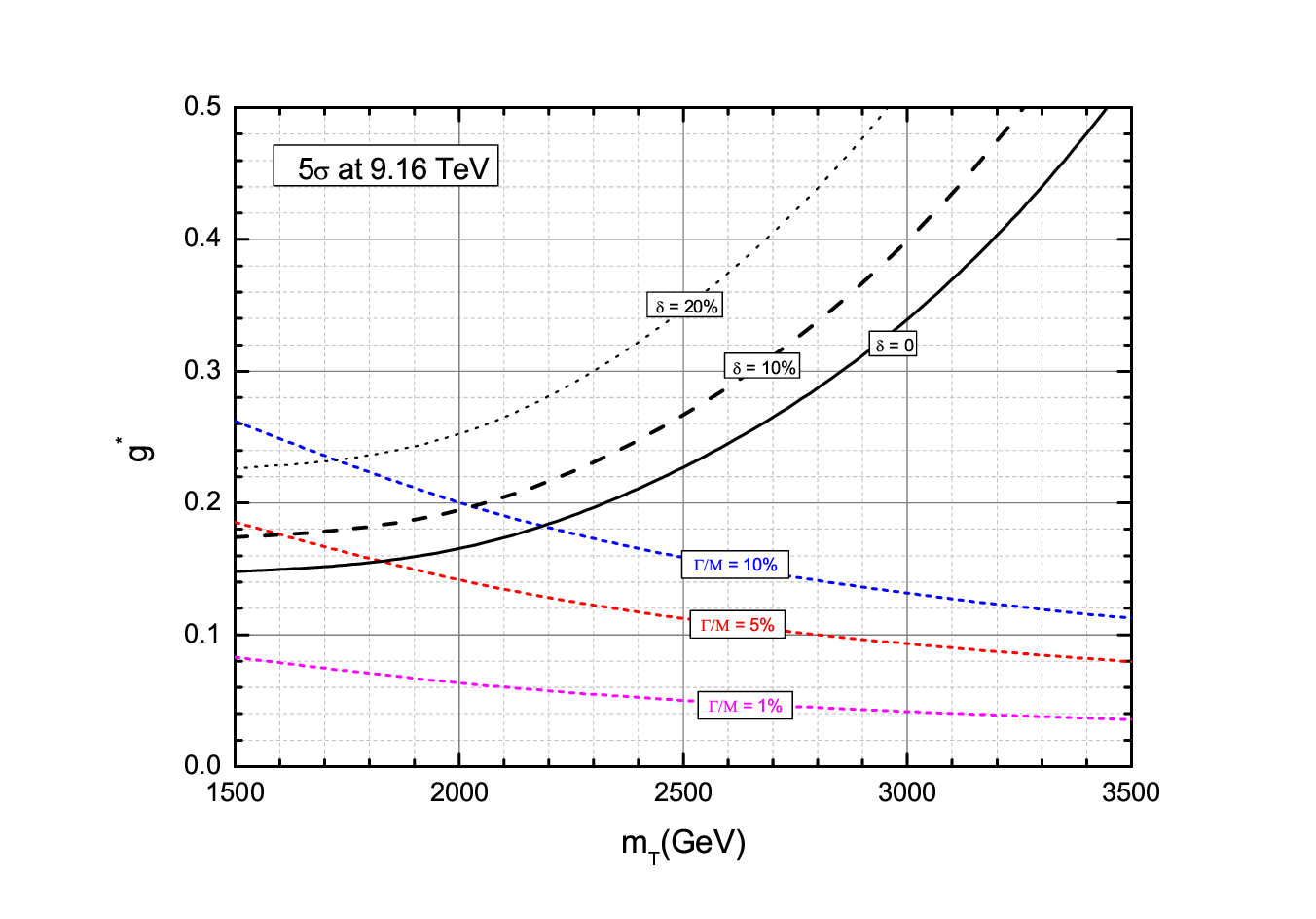}}
\caption{95\% CL exclusion limit (left panel) and $5\sigma$ discovery reach (right panel) contour plots for the signal in $g^{*}-m_T$ in the $W_{\text{lep}}$ channel at the three different c.m. energies. Short-dashed lines denote the contours of $\Gamma_{T}/m_{T}$. }
\label{fig6}
\end{center}
\end{figure}

\begin{figure}[htb]
\begin{center}
\vspace{1.5cm}
\centerline{\epsfxsize=6.5cm \epsffile{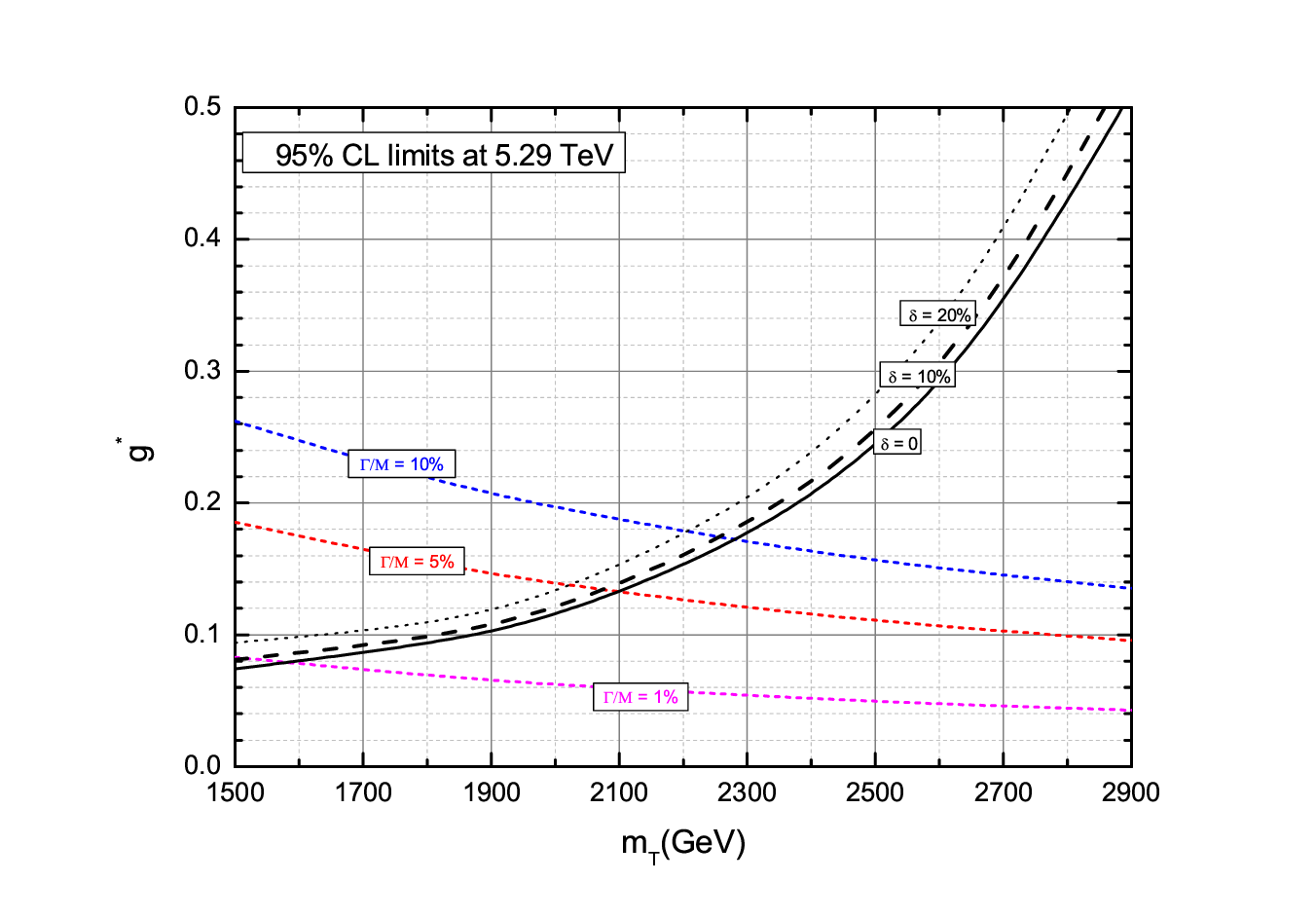}\epsfxsize=6.5cm \epsffile{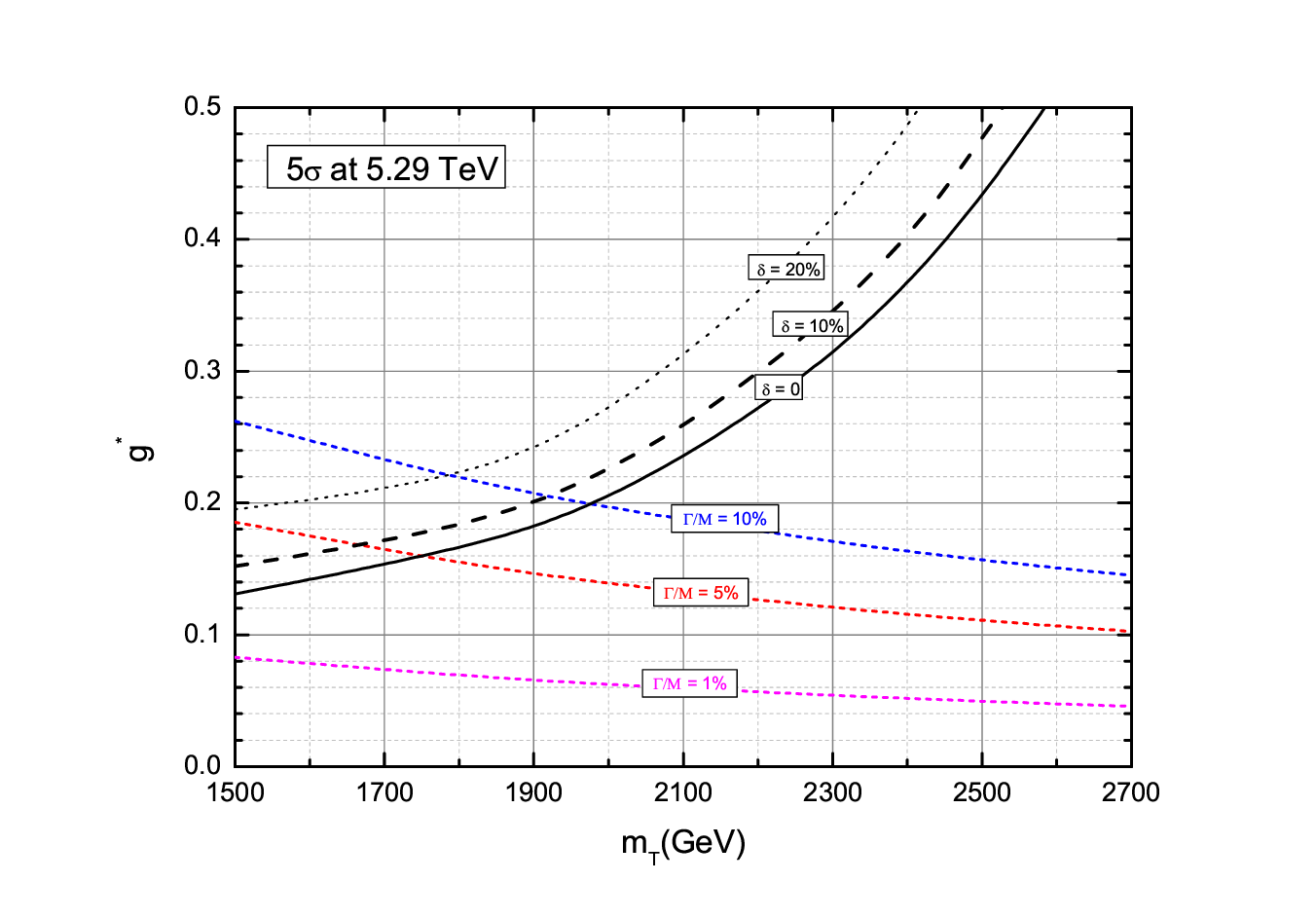}}
\centerline{\epsfxsize=6.5cm \epsffile{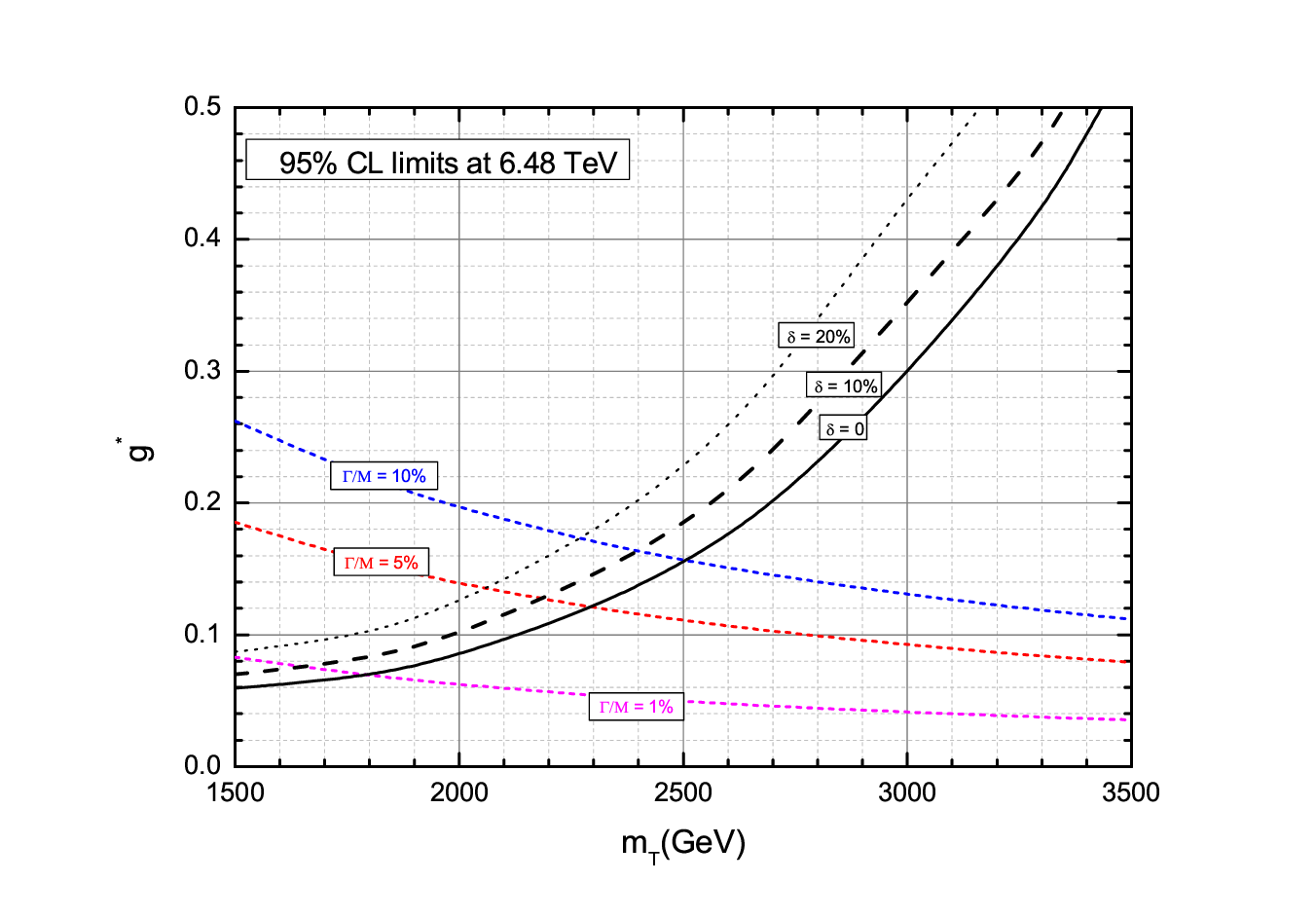}\epsfxsize=6.5cm \epsffile{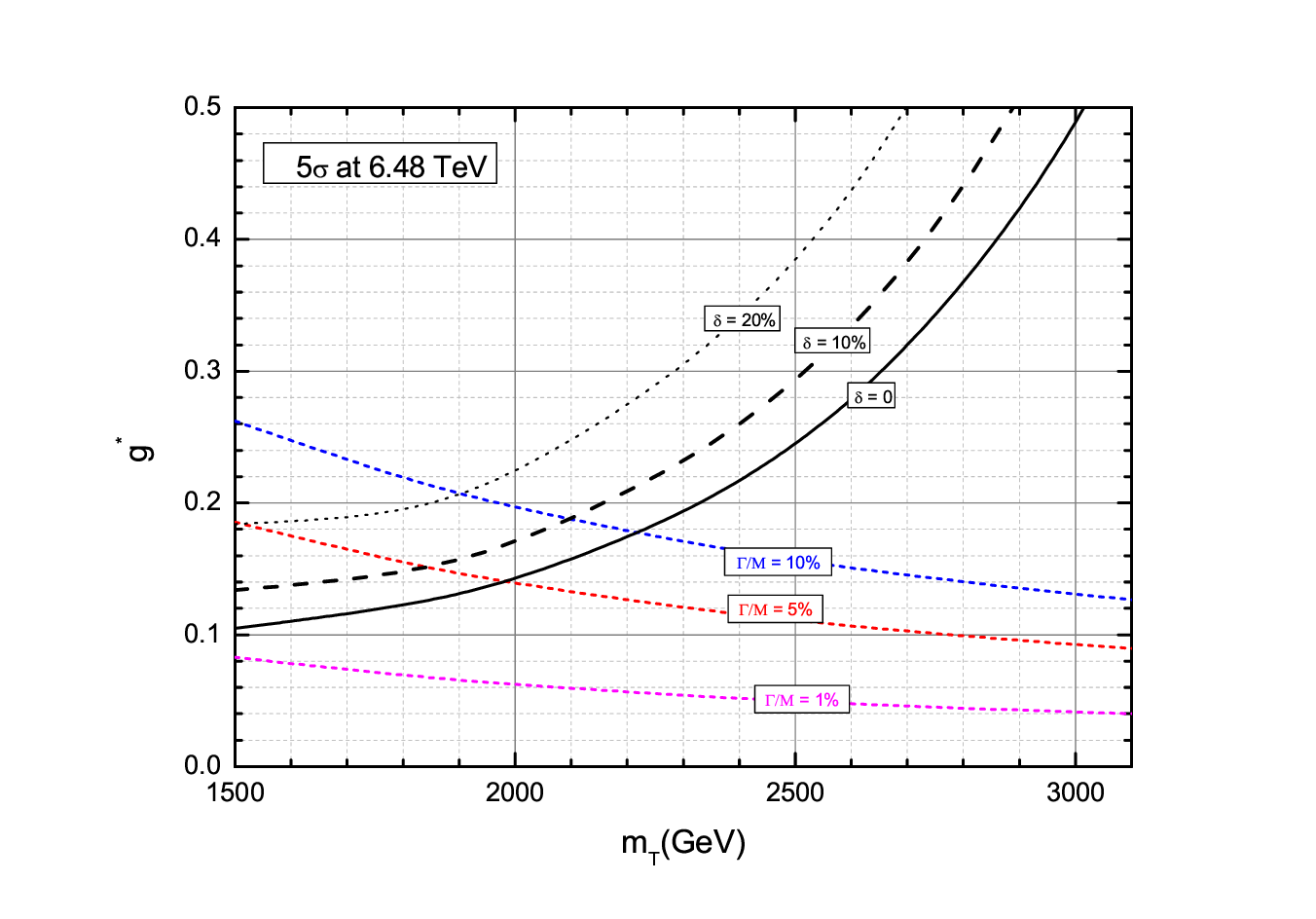}}
\centerline{\epsfxsize=6.5cm \epsffile{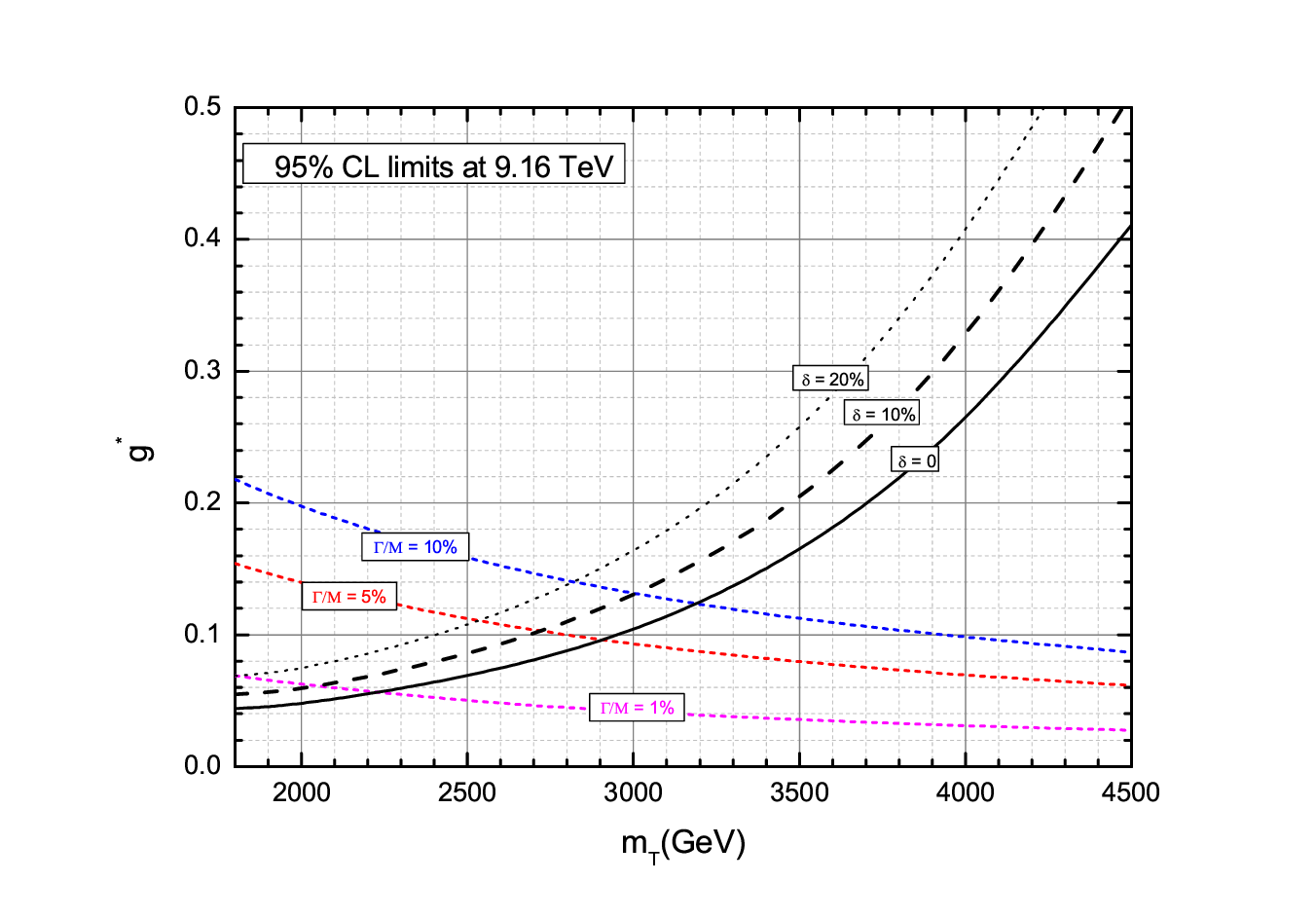}\epsfxsize=6.5cm \epsffile{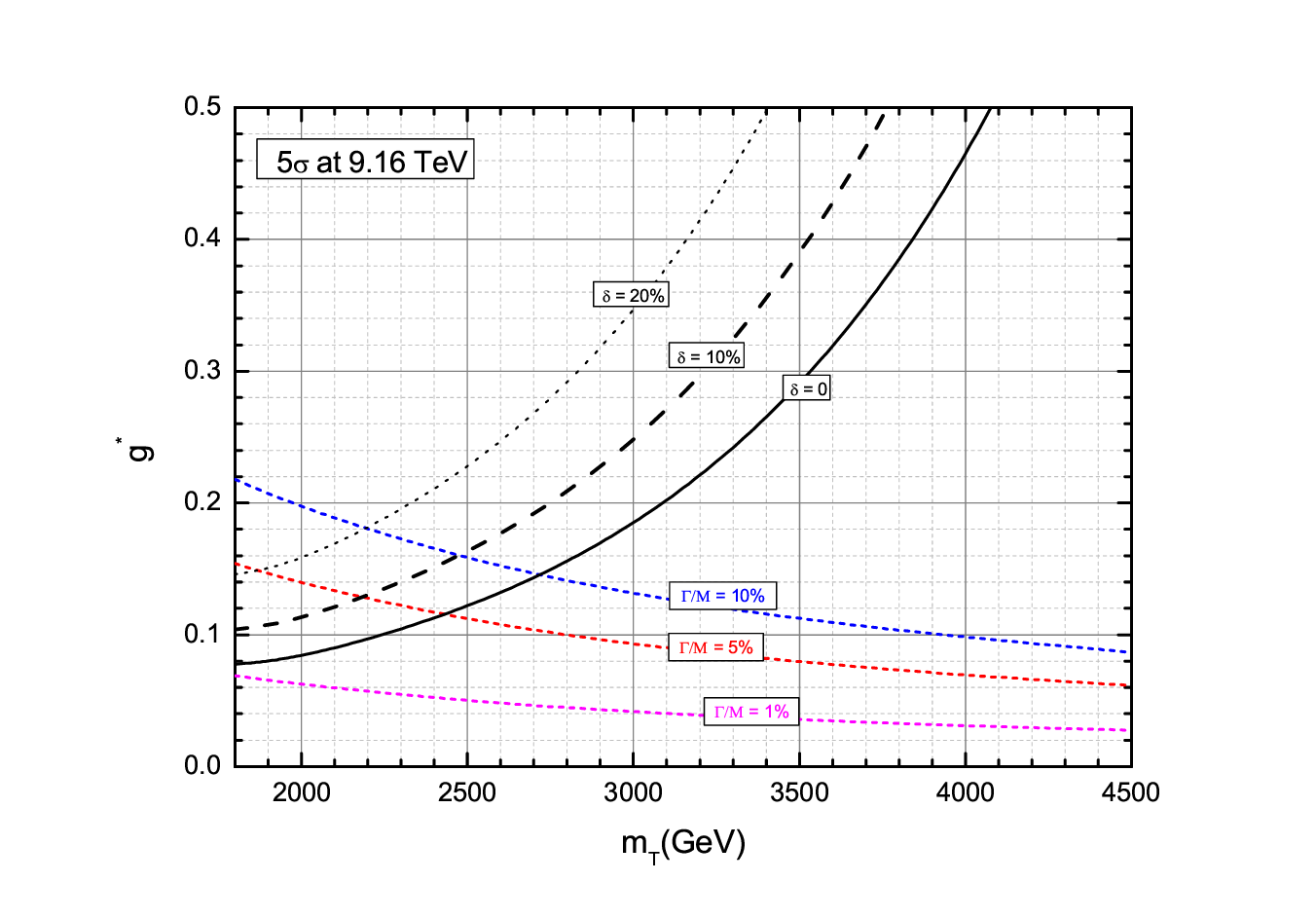}}
\caption{95\% CL exclusion limit (left panel) and $5\sigma$ discovery reach (right panel) contour plots for the signal in $g^{*}-m_T$ in the $W_{\text{had}}$ channel at the three different c.m. energies. Short-dashed lines denote the contours of $\Gamma_{T}/m_{T}$. }
\label{fig7}
\end{center}
\end{figure}

As a result, in Fig.~\ref{fig6}, we present the  95\% CL exclusion limit and $5\sigma$ discovery reaches in the plane of $g^{*}-m_T$ at the three different c.m. energies in the $W_{\text{lep}}$ channel. To illustrate the effect of systematic uncertainty on the significance, we select three cases: no systematics ($\delta=0$) and two typical systematic
uncertainties~($\delta=10\%$ and $\delta=20\%$).  One can see that the higher systematic uncertainty of background can decrease the discovery capability and the exclusion power. With a realistic 10\% systematic error, the discoverable (at $5\sigma$ level) region is  $g^{*}\in [0.3, 0.5]$ with $m_{T}\in [1500, 2100]$~GeV at $\sqrt{s}= 5.29$~TeV,  $g^{*}\in [0.27, 0.5]$ with $m_{T}\in [1500, 2400]$~GeV at $\sqrt{s}= 6.48$~TeV, and $g^{*}\in [0.18,0.5]$ with $m_{T}\in [1500, 3250]$~GeV at $\sqrt{s}= 9.16$~TeV. As for the exclusion (at 95\% CL) region, this is $g^{*}\in [0.17, 0.5]$ with $m_{T}\in [1500, 2500]$~GeV at $\sqrt{s}= 5.29$~TeV, $g^{*}\in [0.14, 0.5]$ with $m_{T}\in [1500, 2900]$~GeV at $\sqrt{s}= 6.48$~TeV,  and $g^{*}\in [0.09, 0.5]$ with $m_{T}\in [1500, 3700]$~GeV at $\sqrt{s}= 9.16$~TeV.

In Fig.~\ref{fig7}, we present the 95\% CL exclusion limit and $5\sigma$ discovery reach in the plane of $g^{*}-m_T$ at the three different c.m. energies in the $W_{\text{had}}$ channel. One finds that the sensitivities are better than those for the  signal in the $W_{\text{lep}}$ channel.
In the presence of 10\% systematic uncertainty, the discoverable (at $5\sigma$ level) region is  $g^{*}\in [0.15, 0.5]$ with $m_{T}\in [1500, 2520]$~GeV at $\sqrt{s}= 5.29$~TeV, $g^{*}\in [0.14, 0.5]$ with $m_{T}\in [1500, 2900]$~GeV at $\sqrt{s}= 6.48$~TeV, and $g^{*}\in [0.1,0.5]$ with $m_{T}\in [1800, 3750]$~GeV at $\sqrt{s}= 9.16$~TeV. As for the exclusion (at 95\% CL) region, this is $g^{*}\in [0.08, 0.5]$ with $m_{T}\in [1500, 2850]$~GeV at $\sqrt{s}= 5.29$~TeV, $g^{*}\in [0.07, 0.5]$ with $m_{T}\in [1500, 3340]$~GeV at $\sqrt{s}= 6.48$~TeV,  and $g^{*}\in [0.06, 0.5]$ with $m_{T}\in [1800, 4500]$~GeV at $\sqrt{s}= 9.16$~TeV.

So far, we have treated the VLQ-$T$ signals in narrow width approximation
(NWA), which is appropriate when the width-to-mass ratio $\Gamma/M$ of the resonance is per mille level, however, for values at percent level and beyond, the full Breit-Wigner (BW) should be adopted, as demonstrated in~\cite{Banerjee:2024zvg}. For the high-coupling scenarios where $\Gamma/m$ becomes substantial, a full off-shell calculation including interference would be necessary, making our NWA results conservative upper bounds. Future studies will incorporate these effects through complete matrix-element calculations. In order to appreciate the difference between the
two approaches in our case, we proceed as in Ref.~\cite{Barducci:2013zaa} and limit ourselves to sample $\Gamma_T/m_T$ values below
10\%. For large mass values of the VLQ-$T$, the decay width is approximated as $\Gamma_{T}\sim (g^{*})^{2}m_{T}^{3}\times 6.5\times10^{-7}$~GeV$^{-2}$, and therefore $g^{*}$ can be chosen to obtain a specific $\Gamma_{T}/m_{T}$ ratio.
Thus
 in Figs.~\ref{fig6} and ~\ref{fig7}, we also display the contours of $\Gamma_{T}/m_{T}$  for three typical values: 1\%, 5\%, and 10\%.
For a fixed value of $\Gamma_{T}/m_{T}=10\%$ and systematic uncertainty of  10\% within, e.g., the $W_{\text{had}}$ channel (the dominant final state),  the signal can be discovered (at $5\sigma$ level) for a VLQ-$T$ mass of about 1900, 2000, and 2500~GeV at  $\sqrt{s}= 5.29,~6.48$, and 9.16~TeV, respectively. For the case of exclusion (at 95\% CL),  the lower limit on the VLQ-$T$ mass is about 2250, 2400, and 3000~GeV, at  $\sqrt{s}= 5.29,~6.48$, and 9.16~TeV, respectively. Hence, width effects can generally be significant.

\begin{figure}[htb]
\begin{center}
\vspace{1.5cm}
\centerline{\epsfxsize=8cm \epsffile{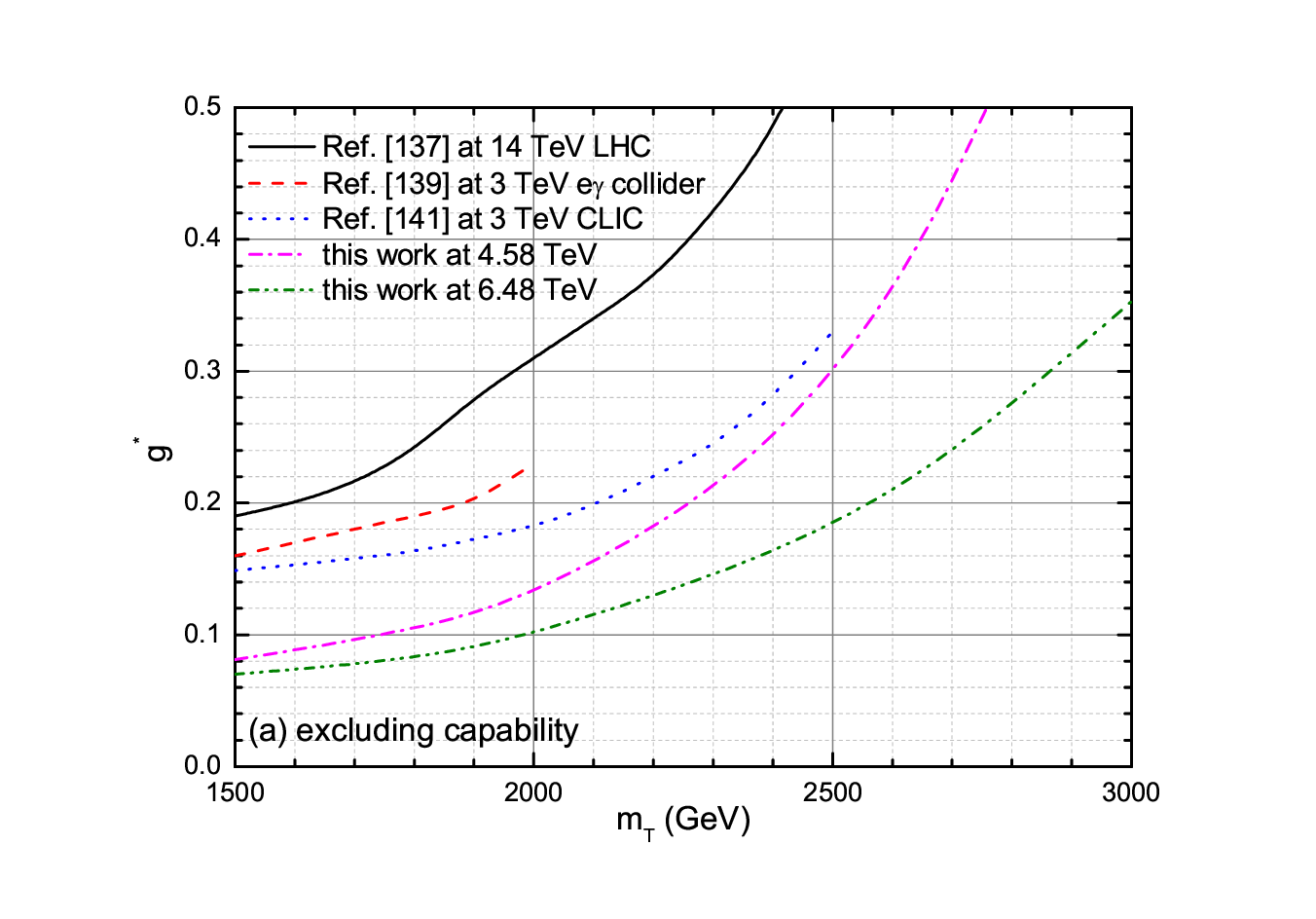}\epsfxsize=8cm \epsffile{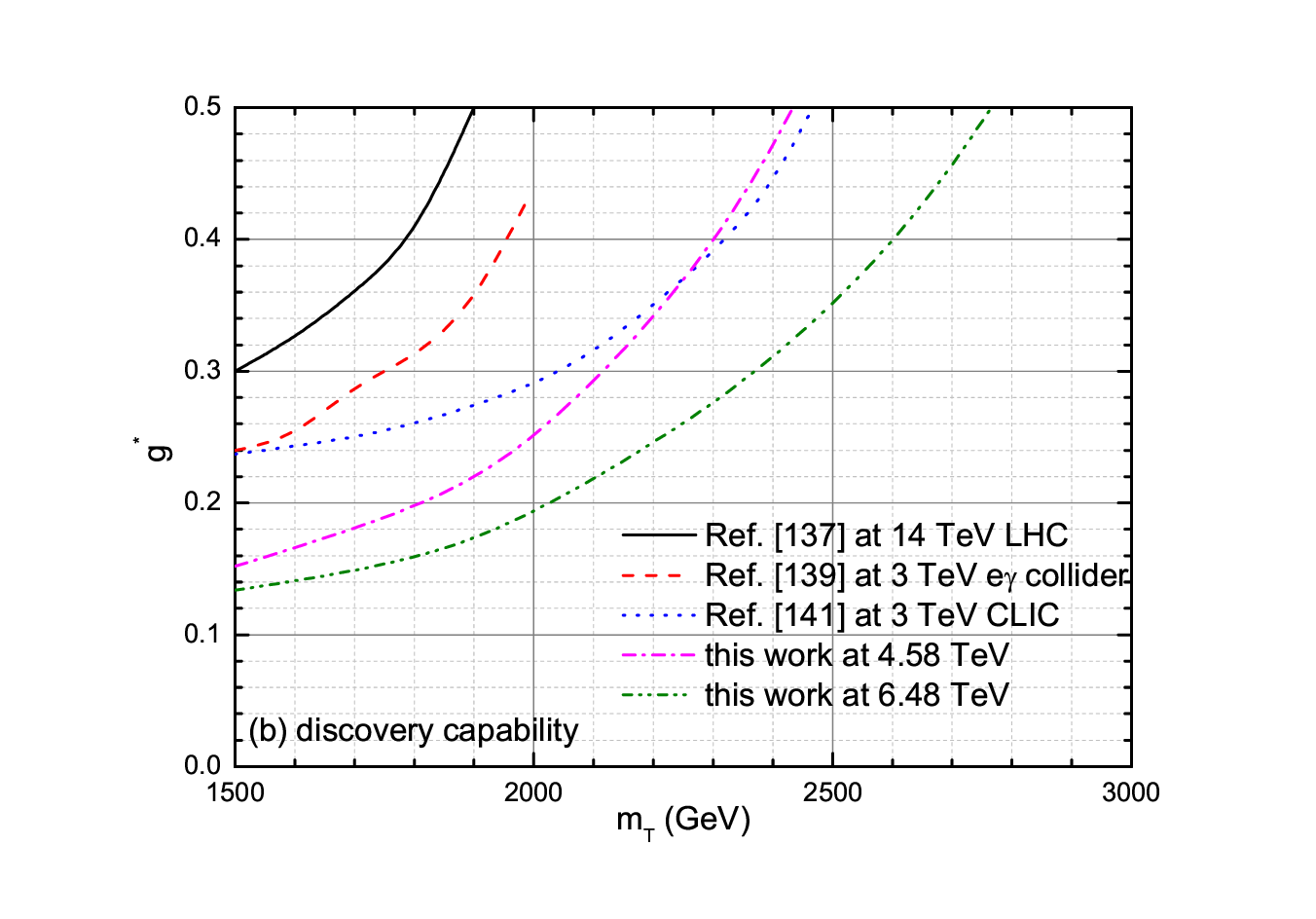}}
\caption{(a) Exclusion limits and (b) discovery reaches for the singlet VLQ-$T$ in the $g^\ast$-$m_T$ plane, comparing sensitivities across future high-energy colliders. Results include projections for LHC (14 TeV, 3 ab$^{-1}$), $e\gamma$ (3 TeV, 3 ab$^{-1}$), $e^+e^-$ (3 TeV, 5 ab$^{-1}$), and $\mu p$ (4.58 and 9.16 TeV, 100 fb$^{-1}$) colliders from Refs.~\cite{Yang:2021btv, Shang:2019zhh, Qin:2022mru}. }
\label{fig8}
\end{center}
\end{figure}

Very recently, the authors of Ref.~\cite{Benbrik:2024fku} have presented a comprehensive review of the most up-to-date exclusion limits
on VLQs derived from ATLAS and CMS data at the LHC, wherein single VLQ-$T$  production constrains
the mixing parameter $\kappa$ to values below 0.26~(0.42) for $m_{T}\sim1.5~(2.0)$ TeV. To enable more intuitive comparison of VLQ-$T$ search capabilities across different collider scenarios, we have supplemented Table~\ref{list} with two contour plots in Fig.~\ref{fig8}, visually presenting the exclusion limits and discovery reaches in the $g^{\ast}-m_T$ parameter space. Our analysis for the $\mu p$ collider assumes a conservative 10\% systematic uncertainty, while other studies employ different assumptions (see respective references for details). This comprehensive comparison demonstrates that a future muon-proton collider would provide complementary sensitivity for singlet VLQ-$T$ searches, particularly in the high-mass region.
\begin{table}[htbp]
\begin{center}
 \caption{\label{list}
Some results of searching for the singlet VLQ-$T$ at different high-energy colliders. Here, the symbol ``$\setminus$" stands for no relevant results in the reference. The results in this work correspond to a mild systematic uncertainty of 10\% at  a $\mu p$ collider with an integrated luminosity of 100 fb$^{-1}$.}
\vspace{0.2cm}
\begin{tabular}{c|c|cc|cc|c}
\hline\hline
\multirow{2}{*}{Channel}      &\multirow{2}{*}{Data Set} &\multicolumn{2}{c|}{Excluding capability} & \multicolumn{2}{c|}{Discovery capability}&\multirow{2}{*}{Reference}  \\
\cline{3-4} \cline{5-6}
&&$g^{\ast}$&$m_T/\rm TeV$&$g^{\ast}$&$m_T/\rm TeV$&  \\
\hline
$T\to tZ$&LHC @14 TeV, 3 ab$^{-1}$&[0.06, 0.25] & [0.9, 1.5]&[0.10, 0.42]&[0.9, 1.5]&\cite{Liu:2017sdg}  \\
$T\to th$&LHC @14 TeV, 3 ab$^{-1}$&[0.16, 0.5] & [1.0, 1.6]&[0.24, 0.72]&[1.0, 1.6]&\cite{Liu:2019jgp}  \\
$T\to bW^{+}$&LHC @14 TeV, 3 ab$^{-1}$&[0.19, 0.5] & [1.3, 2.4]&[0.31, 0.5]&[1.3, 1.9]&\cite{Yang:2021btv}  \\
$T\to bW^{+}$&$e\gamma$ collider @2 TeV, 1 ab$^{-1}$&[0.13, 0.5] & [0.8, 1.6]&$\setminus$&$\backslash$&\cite{Yang:2018fcx}  \\
$T\to tZ$&$e\gamma$ collider @3 TeV, 3 ab$^{-1}$&[0.15, 0.23] & [1.3, 2.0]&[0.23, 0.5] &[1.3, 2.0]&\cite{Shang:2019zhh}  \\
$T\to th$&$e\gamma$ collider @3 TeV, 3 ab$^{-1}$&[0.14, 0.50] & [1.3, 2.0]&[0.27, 0.5] &[1.3, 2.0]&\cite{Shang:2020clm}  \\
$T\to bW^{+}$&$e^{+}e^{-}$ collider @3 TeV, 5 ab$^{-1}$&[0.15, 0.40] & [1.5, 2.6]&[0.24, 0.44] &[1.5, 2.4]&\cite{Qin:2022mru}  \\
 $T\to tZ$&$e^{+}e^{-}$ collider @3 TeV, 5 ab$^{-1}$&[0.19, 0.40] & [1.3, 2.5]&[0.31, 0.5] &[1.3, 2.3]&\cite{Han:2022exz} \\  \hline
\multirow{3}{*}{$T\to bW\to b\ell \nu$}&$\mu p$ collider @4.58 TeV&[0.16, 0.48] & [1.5, 2.5]&[0.31, 0.5] &[1.5, 2.1]&\multirow{6}{*}{This work}  \\
&$\mu p$ collider @6.48 TeV&[0.14, 0.5] & [1.5, 2.9]&[0.24, 0.5] &[1.5, 2.4]& \\
&$\mu p$ collider @9.16 TeV&[0.09, 0.4] & [1.5, 3.7]&[0.18, 0.5] &[1.5, 3.2]&  \\ \cline{1-6}
\multirow{3}{*}{$T\to bW\to bJ$}&$\mu p$ collider @4.58 TeV&[0.08, 0.5] & [1.5, 2.85]&[0.14, 0.5] &[1.5, 2.52]& \\
&$\mu p$ collider @6.48 TeV&[0.07, 0.5] & [1.5, 3.34]&[0.13, 0.5] &[1.5, 2.9]& \\
&$\mu p$ collider @9.16 TeV&[0.06, 0.5] & [2.0, 4.5]&[0.1, 0.5] &[1.8, 3.75]&\\
\hline
 \end{tabular}
 \end{center}
\end{table}

\subsection{Expected discovery and exclusion reaches for VLQ-$Y$}
\begin{figure}[t]
\begin{center}
\vspace{1.5cm}
\centerline{\epsfxsize=6.5cm \epsffile{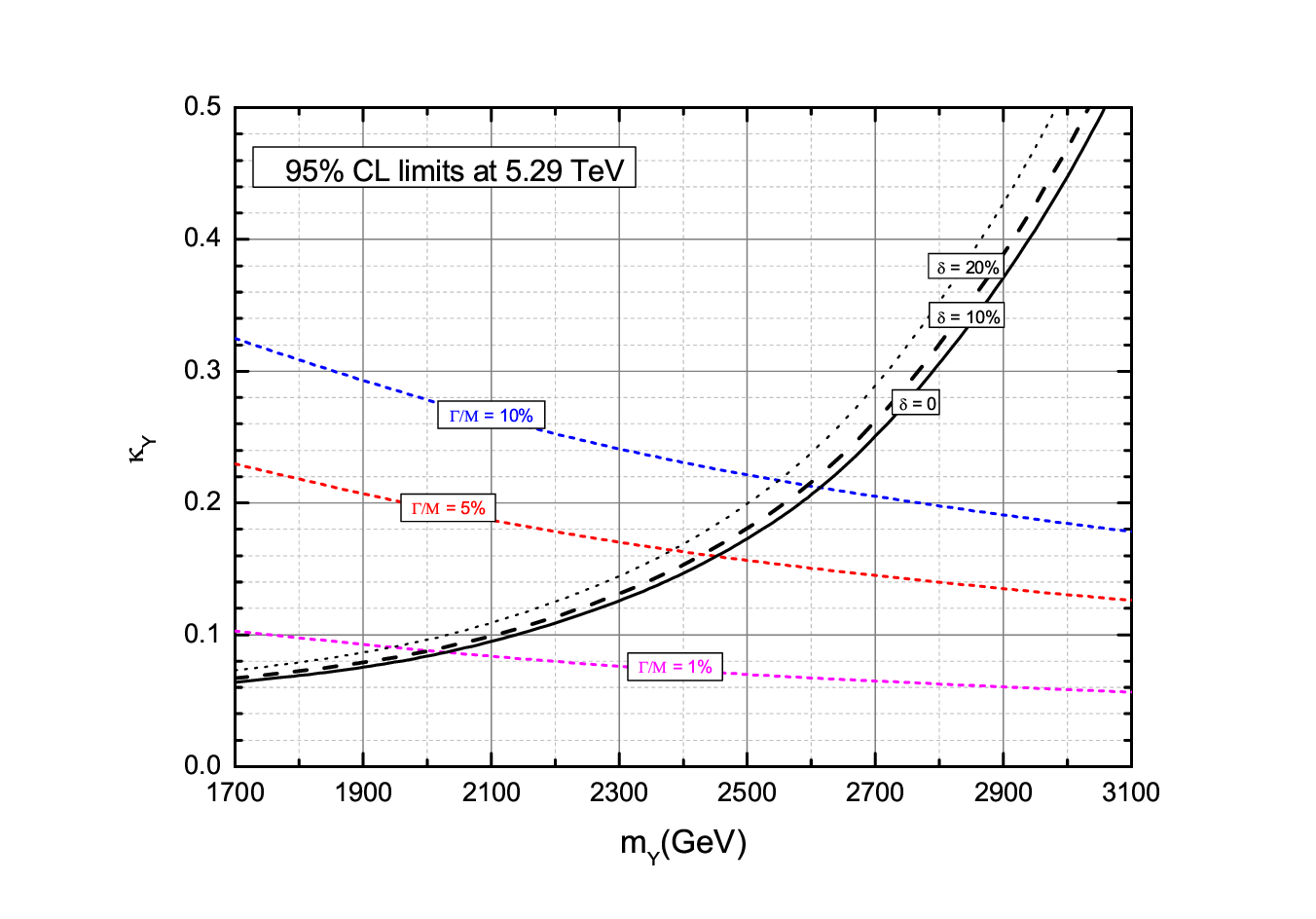}\epsfxsize=6.5cm \epsffile{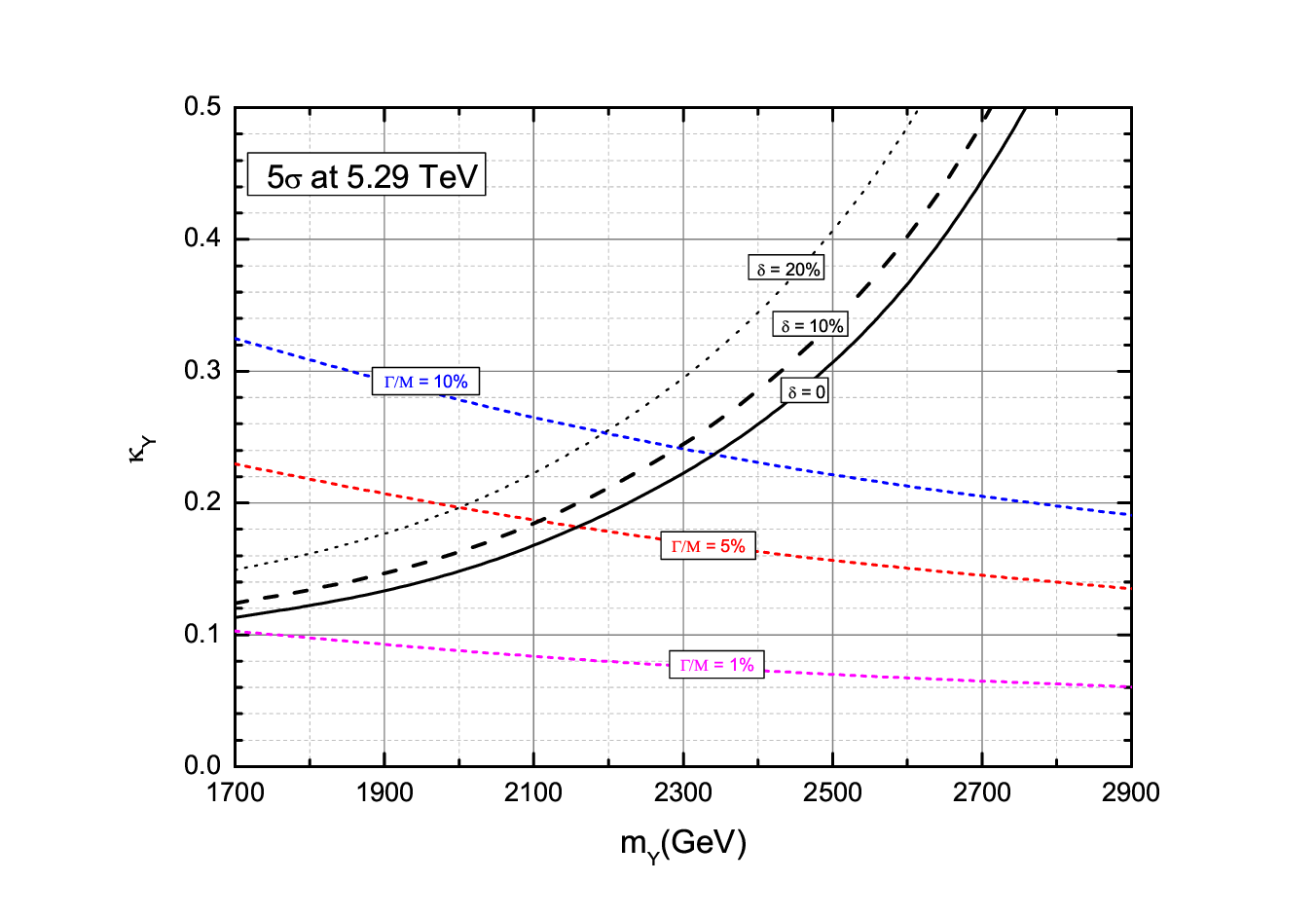}}
\centerline{\epsfxsize=6.5cm \epsffile{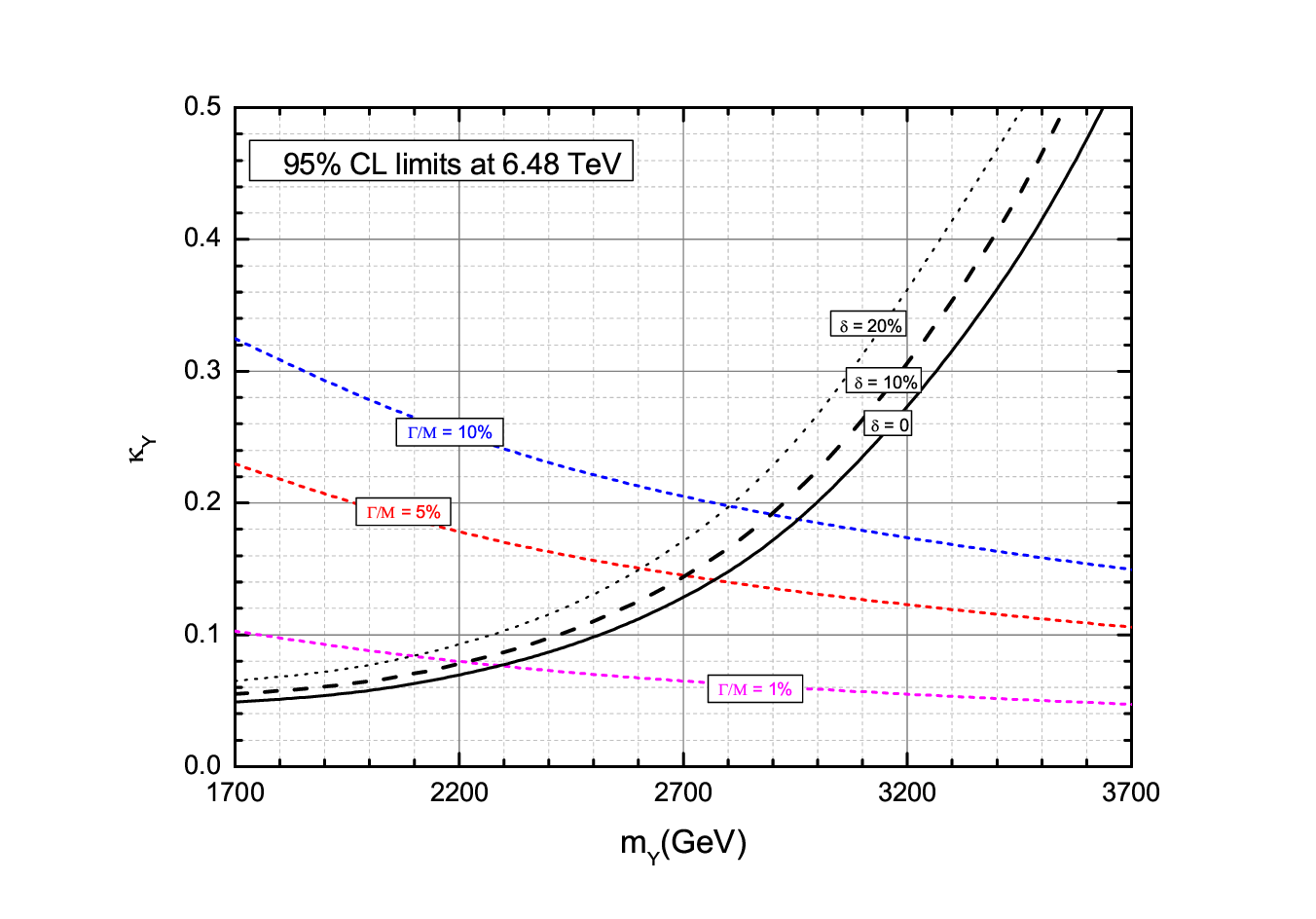}\epsfxsize=6.5cm \epsffile{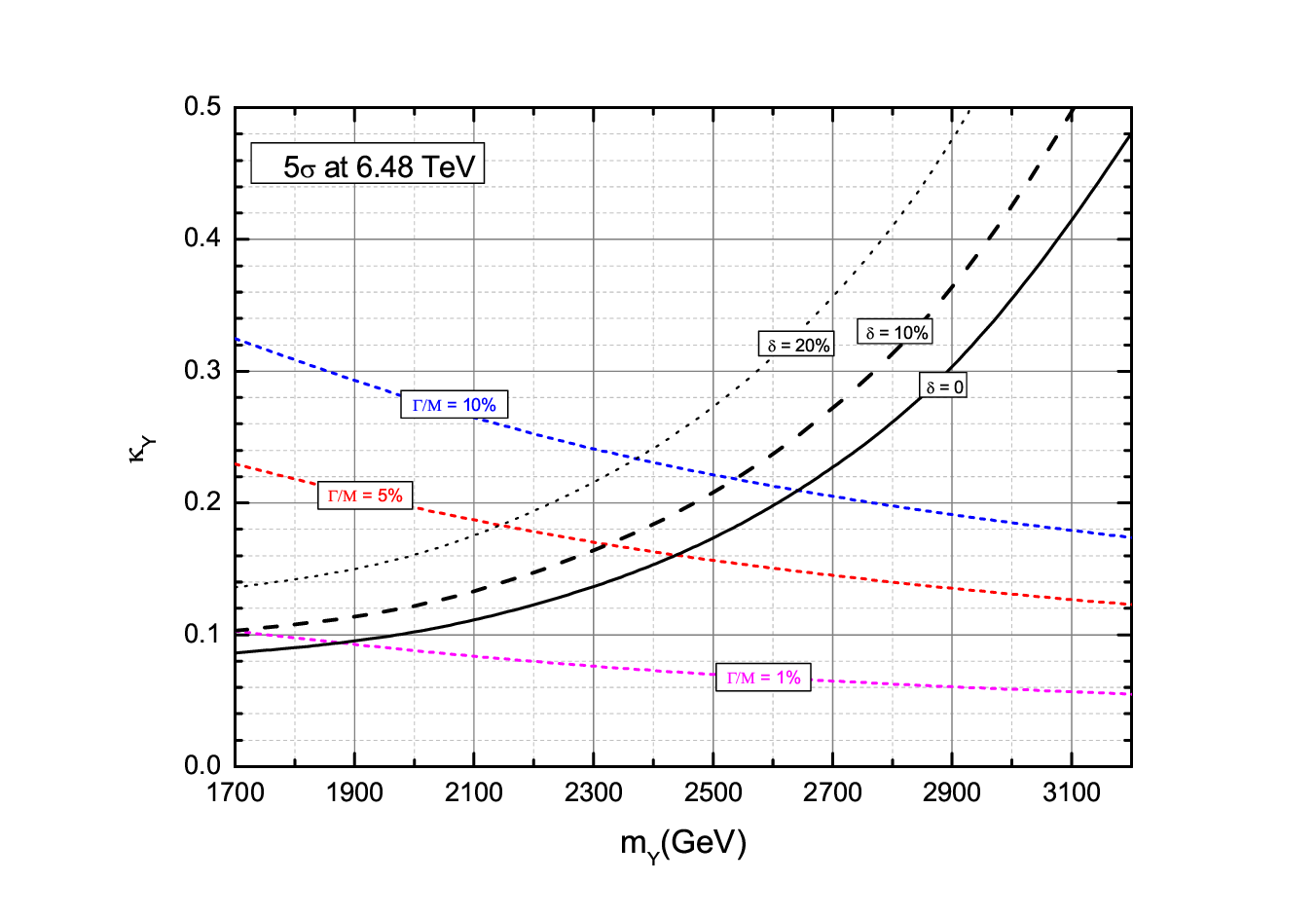}}
\centerline{\epsfxsize=6.5cm \epsffile{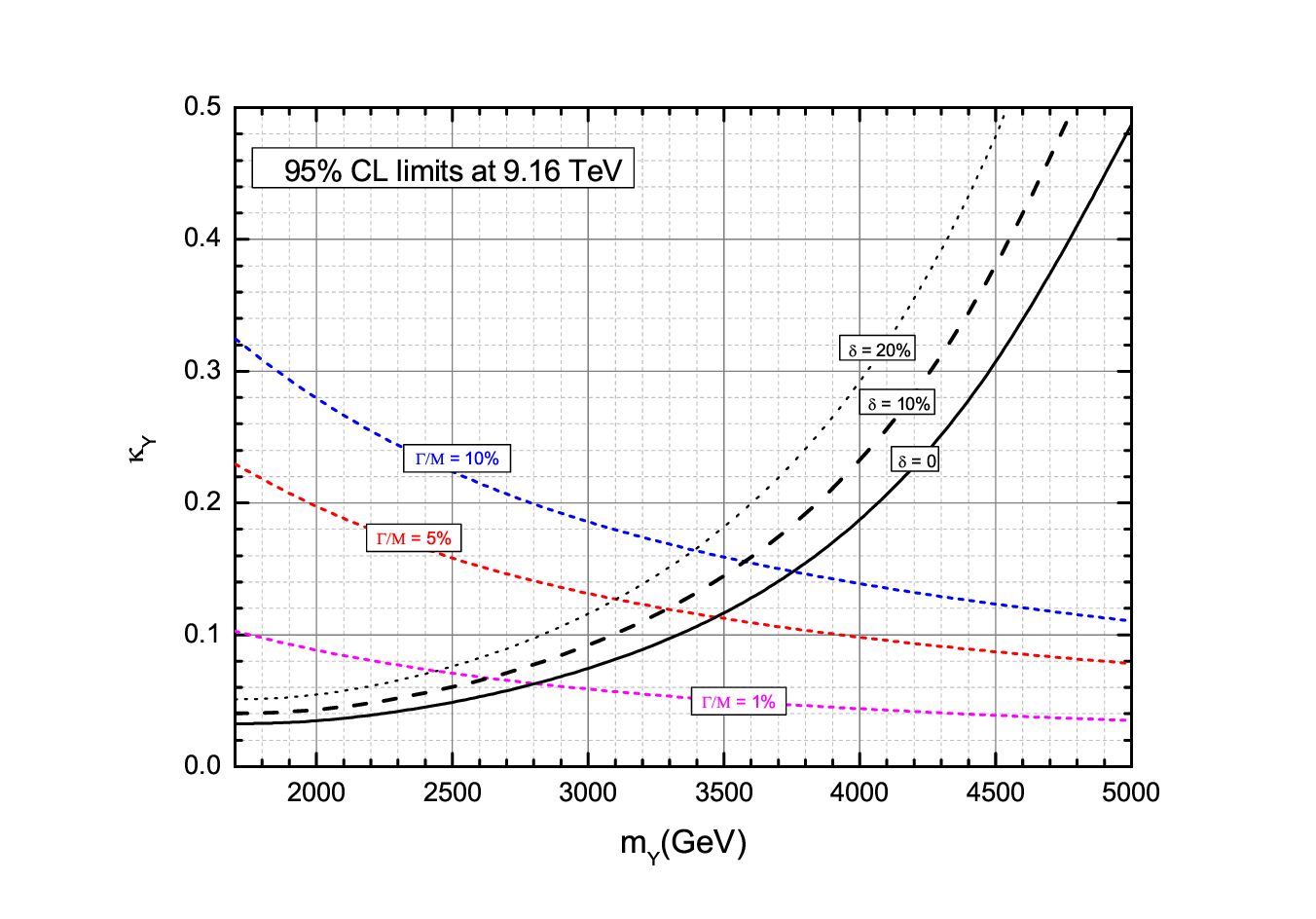}\epsfxsize=6.5cm \epsffile{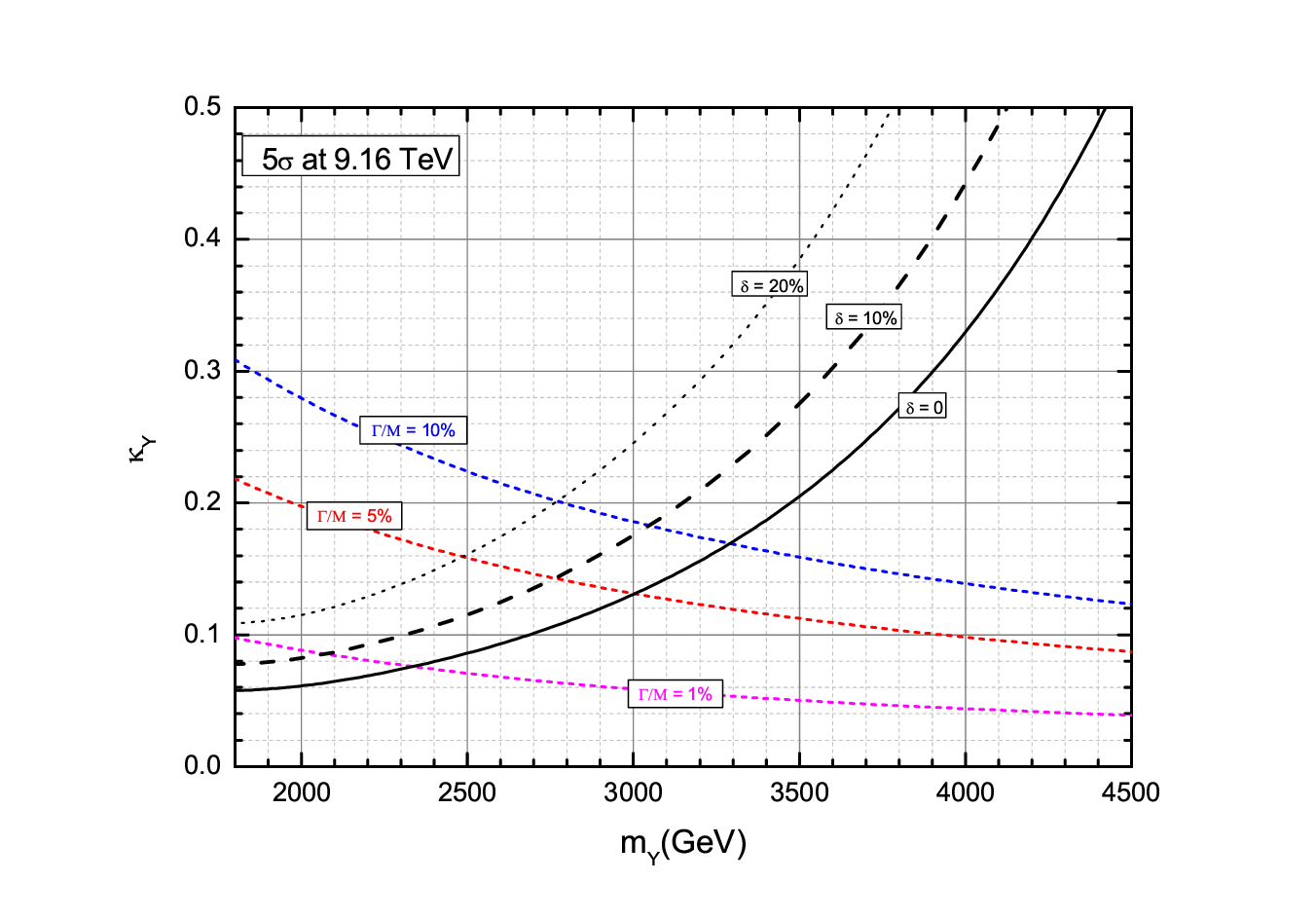}}
\caption{95\% CL exclusion limit (left panel) and $5\sigma$ discovery reach (right panel) contour plots for the signal in $\kappa-m_Y$ at the three different center-of-mass energies. Short-dashed lines denote the contours of $\Gamma_{Y}/m_{Y}$. }
\label{fig9}
\end{center}
\end{figure}
As explained, the MC analysis for the VLQ-$Y$ case is identical to the previous one. Thus, similarly to what previously done, in Fig.~\ref{fig9}, we present the  95\% CL exclusion limit and $5\sigma$ discovery reach in the plane of $\kappa_{Y}-m_Y$ at the usual three different c.m. energies, albeit limitedly to the fat jet final state (the dominant one, thereby neglecting the subdominant leptonic signal).
In the presence of 10\% systematic uncertainty, the discoverable (at $5\sigma$ level) region is  $\kappa_{Y}\in [0.12, 0.5]$ with $m_{Y}\in [1700, 2700]$~GeV at $\sqrt{s}= 5.29$~TeV, $\kappa_{Y}\in [0.10, 0.5]$ with $m_{Y}\in [1700, 3100]$~GeV at $\sqrt{s}= 6.48$~TeV, and $\kappa_{Y}\in [0.08,0.5]$ with $m_{Y}\in [1700, 4100]$~GeV at $\sqrt{s}= 9.16$~TeV. Instead, a VLQ-$Y$ can be excluded (at 95\% CL)  for $\kappa_{Y}\in [0.07, 0.5]$ with $m_{Y}\in [1700, 3020]$~GeV at $\sqrt{s}= 5.29$~TeV, $\kappa_{Y}\in [0.06, 0.5]$ with $m_{Y}\in [1700, 3540]$~GeV at $\sqrt{s}=6.48$~TeV,  and $\kappa_{Y}\in [0.04, 0.5]$ with $m_{Y}\in [1700, 4800]$~GeV at $\sqrt{s}= 9.16$~TeV.
Furthermore, for a fixed value of $\Gamma_{Y}/m_{Y}=10\%$,  the VLQ-$Y$ can be discovered (at $5\sigma$ level) with a mass of about 2300, 2500, and 3000~GeV at  $\sqrt{s}= 5.29,~6.48$, and 9.16~TeV, respectively. Instead,
the 95\% CL excluded region for the VLQ-$T$ mass is below about 2600, 2900, and 3600~GeV at  $\sqrt{s}= 5.29,~6.48$, and 9.16~TeV, respectively.

Very recently, the authors of Ref.~\cite{Shang:2024wwy} investigated the expected
limits for the VLQ-$Y$ state via single production of $Y$ followed by decay channel $Y\to Wb$ at the LHC with $\sqrt{s}=14$~TeV and the future high-energy $pp$ colliders.
 Considering an integrated luminosity of 300~(3000) fb$^{-1}$ at the 14 TeV LHC with a systematic uncertainty $\delta=10\%$,  the VLQ-$Y$  can be discovered (at $5\sigma$ level) over the region $\kappa_{Y}\in [0.11, 0.5]$ with $m_{Y}\in [1500, 3200]$~GeV~($\kappa_{Y}\in [0.1, 0.5]$ with  $m_{Y}\in [1500, 3350]$~GeV),
and excluded (at 95\% CL) over the  region $\kappa_{Y}\in [0.06, 0.5]$ with $m_{Y}\in [1500, 3800]$~GeV~($\kappa_{Y}\in [0.05, 0.5]$ with  $m_{Y}\in [1500, 3970]$~GeV).
 For a fixed value of $\Gamma_{Y}/m_{Y}=10\%$,  the VLQ-$Y$ can be discovered (excluded) with a mass about 2200~(2600)~GeV at the High-Luminosity
 LHC~(HL-LHC).
 Besides, the authors of Ref.~\cite{Cetinkaya:2020yjf} studied single production of VLQ-$Y$ at the HL-LHC  with $\sqrt{s}=14$~TeV via the fully hadronic mode $Y\to bW\to bjj$. For the integrated luminosity projection of 3000 fb$^{-1}$ and $\kappa_{Y}=0.5~(0.3)$, the lower limits for $m_{Y}$ were obtained as 2350~(1550) GeV for exclusion (at $95\%$ CL), and 1900~(1250) GeV for discovery (at $5\sigma$  level).
 Thus, we conclude again that our study can drive complementary searches for a possible doublet VLQ-$Y$ at a future muon-proton collider.
\section{CONCLUSIONS}
In this paper, we have studied the potential of a future $\mu p$ collider to search for heavy  VLQs of type $T$ and  $Y$  via the single production mode $\mu g\to \nu_\mu b T/Y$ and  subsequent decay $T/Y\to bW$. To be as model-independent as possible, a simplified framework with only two free parameters was applied:
 the VLQs mass $m_{T/Y}$ and  the EW coupling constant $g^{\ast}/\kappa_{Y}$.
Specifically,  we have presented a search strategy at such a possible future machine with the three typical c.m. energies
$\sqrt{s}=5.29$, $6.48$, and $9.16$ TeV,
 for VLQ-$T$ and VLQ-$Y$ signals with two final states: one electrons plus one $b$-tagged jet and missing energy in the $W_{\text{lep}}$ channel, and one fat jet plus one $b$-tagged jet and missing energy in the $W_{\text{had}}$ channel.  After performing a detector level simulation for the signal and relevant SM backgrounds, the $5\sigma$ discovery prospects and $95\%$ CL exclusion limits over  the relevant parameter space were obtained.

From the numerical results, all tested against existing experimental results from the LHC from  searches for VLQ, we have obtained the following results.
\begin{enumerate}
\item
Due to a larger cross section for the final state including a fat $W$-jet with respect to the leptonic $W$ decay channel, the sensitivities in the $W_{\text{had}}$ channel are better than those in the $W_{\text{lep}}$ channel, respectively.

\item
Considering a systematic uncertainty of 10\% with an
integrated luminosity of 100 fb$^{-1}$, for the VLQ-$T$, the discoverable (at $5\sigma$ level) region is  $g^{*}\in [0.15, 0.5]$ with $m_{T}\in [1500, 2520]$~GeV at $\sqrt{s}=5.29$~TeV,  gradually changing to $g^{*}\in [0.13, 0.5]$ with $m_{T}\in [1500, 2900]$~GeV at $\sqrt{s}=6.48$~TeV, and to $g^{*}\in [0.1,0.5]$ with $m_{T}\in [1800, 3750]$~GeV at $\sqrt{s}=9.16$~TeV. Conversely, the 95\% CL exclusion limit is  attained over the
parameter space region $g^{*}\in [0.08, 0.5]$ with $m_{T}\in [1500, 2750]$~GeV at $\sqrt{s}=5.29$~TeV,  $g^{*}\in [0.07, 0.5]$ with $m_{T}\in [1500, 3340]$~GeV at $\sqrt{s}=6.48$~TeV,  and $g^{*}\in [0.06, 0.5]$ with $m_{T}\in [1800, 4500]$~GeV at $\sqrt{s}=9.16$~TeV.

\item
For the above systematic uncertainty and integrated luminosity, in the case of the
VLQ-$Y$, the discoverable (at $5\sigma$ level) region is  $\kappa_{Y}\in [0.12, 0.5]$ with $m_{Y}\in [1700, 2700]$~GeV at $\sqrt{s}=5.29$~TeV, $\kappa_{Y}\in [0.10, 0.5]$ with $m_{Y}\in [1700, 3100]$~GeV at $\sqrt{s}=6.48$~TeV, and $\kappa_{Y}\in [0.08,0.5]$ with $m_{Y}\in [1700, 4100]$~GeV at $\sqrt{s}=9.16$~TeV. Conversely, the  95\% CL exclusion  limit is attained over the
parameter space of $\kappa_{Y}\in [0.07, 0.5]$ with $m_{Y}\in [1700, 3020]$~GeV at $\sqrt{s}=5.29$~TeV,  $\kappa_{Y}\in [0.06, 0.5]$ with $m_{Y}\in [1700, 3540]$~GeV at $\sqrt{s}=6.48$~TeV,  and $\kappa_{Y}\in [0.04, 0.5]$ with $m_{Y}\in [1700, 4800]$~GeV at $\sqrt{s}=9.16$~TeV.

\item
As previous literature has emphasized the role of the $T$ and $Y$ width in the case of their single production mode at the LHC,
for a fixed value of $\Gamma_{T/Y}/m_{T/Y}=10\%$,  the VLQ-$T~(Y)$ can be discovered (at $5\sigma$ level) with a mass about 1900~(2300), 2000~(2500), and 2500~(3000)~GeV at $\sqrt{s}=5.29,~6.48$ and 9.16~TeV, respectively. Conversely,
the 95\% CL exclusion limit is attained for VLQ-$T$ mass about 2250~(2600), 2400~(2900), and 3000~(3600)~GeV, at  $\sqrt{s}=5.29,~6.48$ and 9.16~TeV, respectively.

\end{enumerate}

Finally, by comparing our results to existing literature on the VLQ-$T$ and VLQ-$Y$ states at a variety of present and future colliders, we have concluded that a future $\mu p$ machine can be competitive in the search for such possible new states of Nature.

\begin{acknowledgments}
The work of J-ZH and Y-BL  is supported by the Natural Science Foundation of Henan Province~(Grant No.~242300421398).
The work of SM is supported in part through the NExT Institute and the STFC Consolidated Grant  ST/X000583/1.
\end{acknowledgments}

\begin{appendix}
\section{Event Counts and Selection Efficiencies}\label{appendix}
The cumulative efficiency after a series of selection cuts is defined as the ratio of surviving events ($E_K$) to the initial generated events ($E_{\text{in}}$)
\begin{equation}\label{eff}
\epsilon = \frac{E_K}{E_{\text{in}}}
\end{equation}

\subsection{Analysis at $\sqrt{s}=5.29$~TeV: Leptonic $W$ channel}\label{eff:lep}
The Monte Carlo samples for this channel include:
\begin{itemize}
    \item $10^5$ events for each signal benchmark point: $T_{1500}$, $T_{2000}$, and $T_{2500}$
    \item $5 \times 10^5$ events for each SM background process: $\nu_{\mu}tb$ and $\nu_{\mu}Wj$
\end{itemize}

\begin{figure*}[htb]
\begin{center}
\centerline{\hspace{2.0cm}\epsfxsize=8cm\epsffile{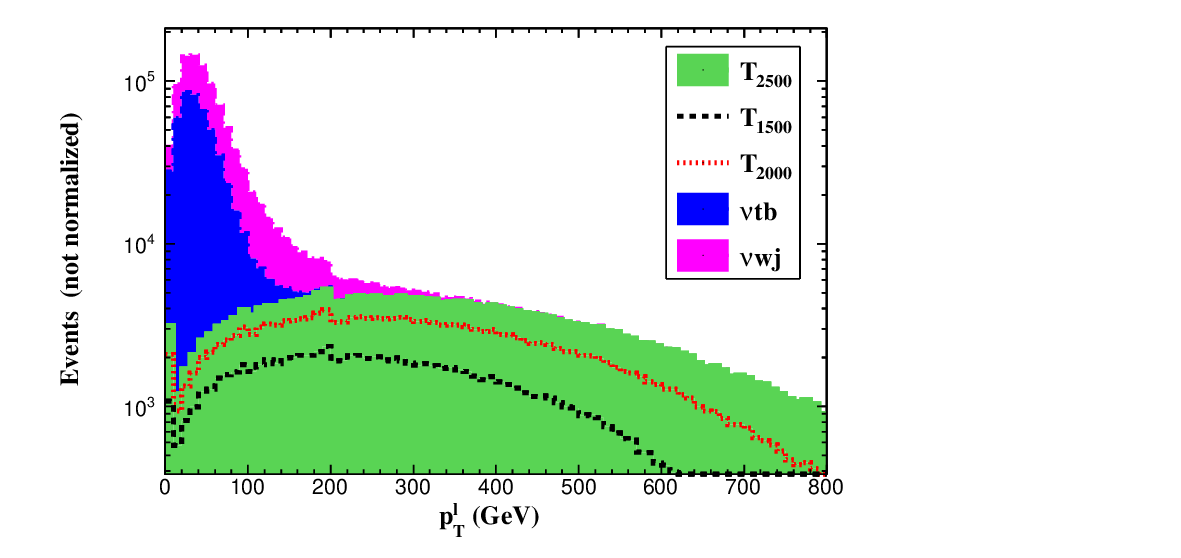}
\hspace{-2.0cm}\epsfxsize=8cm\epsffile{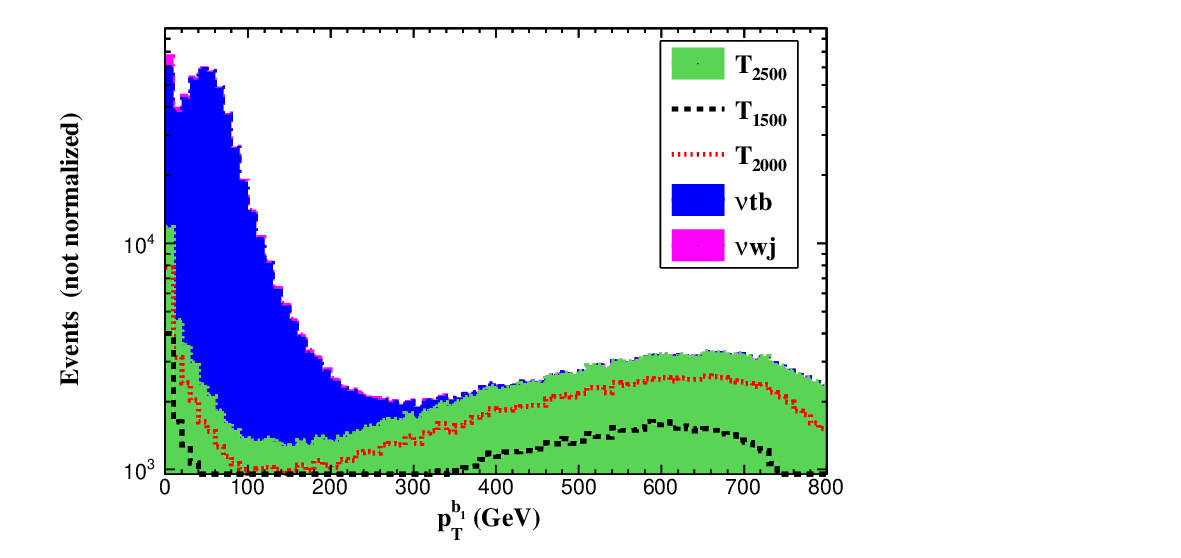}}
\centerline{\hspace{2.0cm}\epsfxsize=8cm\epsffile{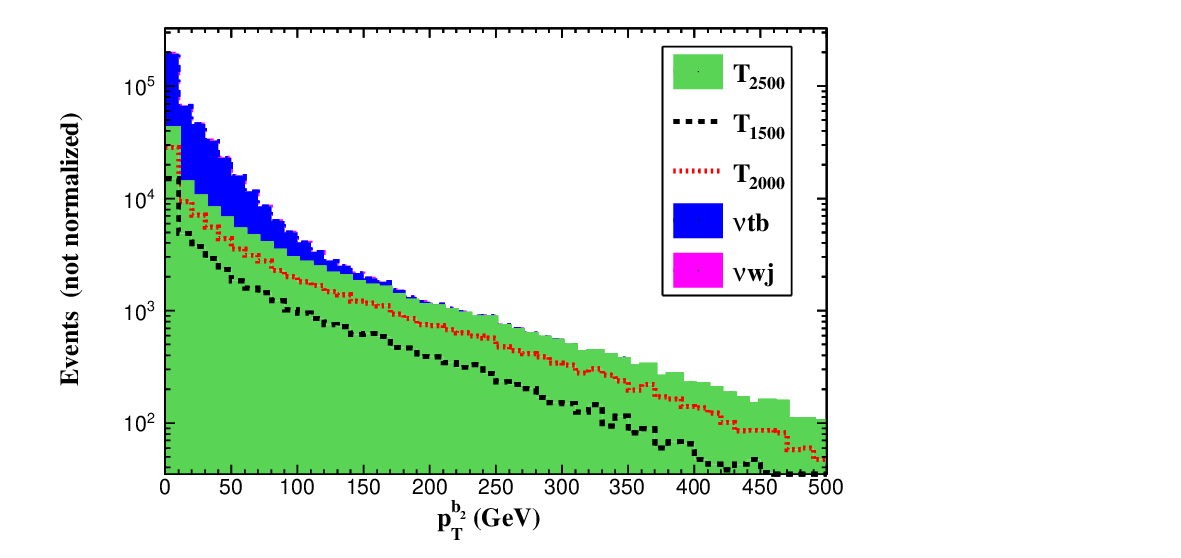}
\hspace{-2.0cm}\epsfxsize=8cm\epsffile{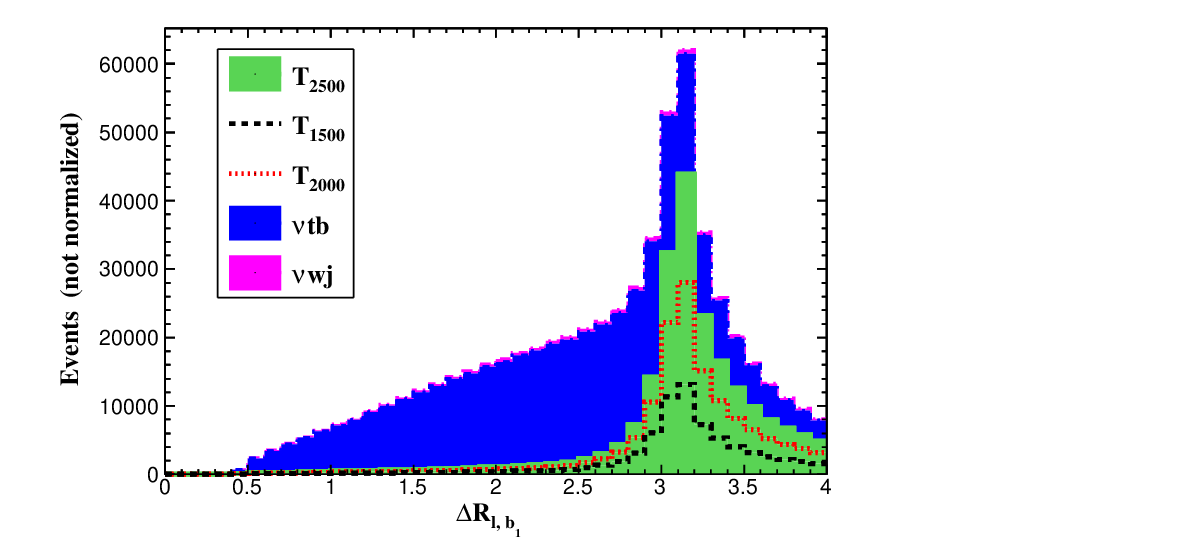}}
\centerline{\hspace{2.0cm}\epsfxsize=8cm\epsffile{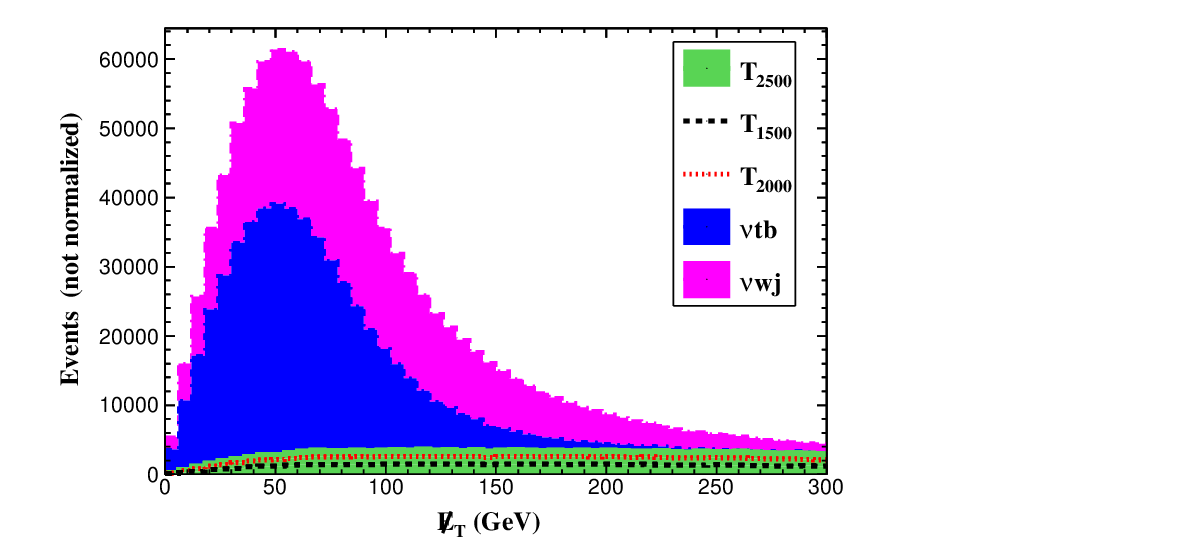}
\hspace{-2.0cm}\epsfxsize=8cm\epsffile{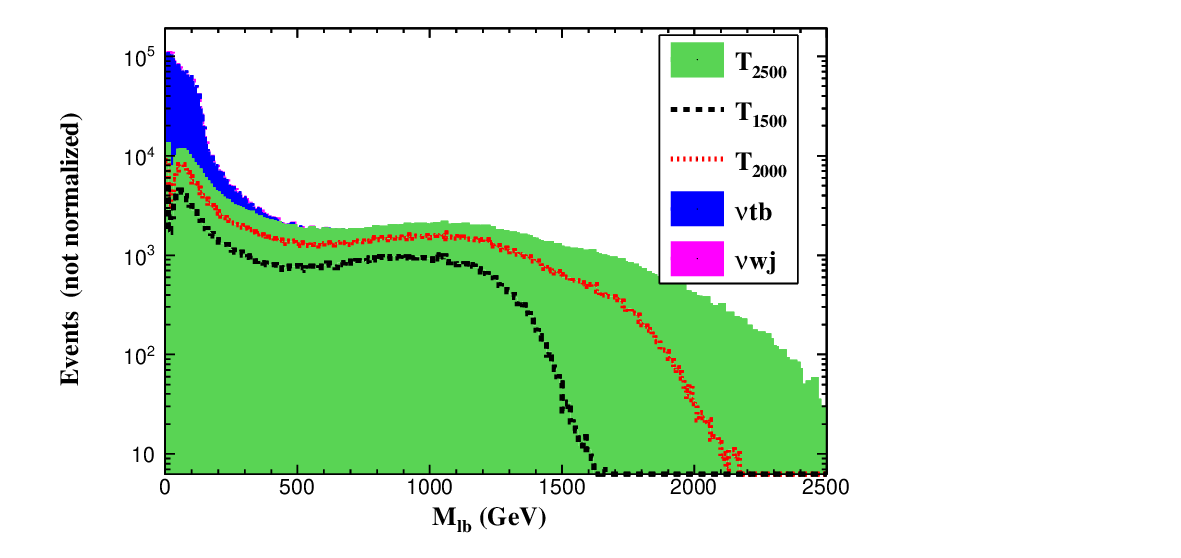}}
\caption{Kinematic distributions prior to selection cuts for $p_T^{\ell}$, $p_T^{b_{1,2}}$, $\Delta R_{\ell,b_1}$, $\slashed{E}_T$, and $M_{be}$ at $\sqrt{s}=5.29$~TeV in the $W_{\text{lep}}$ channel. The green band corresponds to $m_T = 2500$ GeV, while the dashed ($m_T = 1500$ GeV) and dotted ($m_T = 2000$ GeV) lines show independent predictions overlaid for comparison. All distributions display raw Monte Carlo event counts without cross section or luminosity normalization.}
\label{fig10}
\end{center}
\end{figure*}

Figure~\ref{fig10} presents the preselection kinematic distributions for both signal and background processes, showing:
\begin{itemize}
    \item Transverse momentum of the lepton ($p_T^\ell$)
    \item Transverse momenta of the leading and subleading $b$-jets ($p_T^{b_1}$, $p_T^{b_2}$)
    \item Angular separation between the leading $b$-jet and lepton ($\Delta R_{\ell,b_1}$)
    \item Missing transverse energy ($\slashed{E}_T$)
    \item Invariant mass of the $b$-jet and electron system ($M_{be}$)
\end{itemize}

Table~\ref{event1} presents the event event counts and cumulative efficiencies for the leptonic $W$ channel ($W_{\text{lep}}$) at $\sqrt{s} = 5.29$ TeV.

\begin{table}[htb]
\centering
\caption{Event counts and cumulative efficiencies after relevant selection cuts for the $W_{\text{lep}}$ channel at $\sqrt{s}=5.29$~TeV. Values in parentheses show efficiencies\label{event1}}
\begin{tabular}{lcccccc}
\toprule[1.5pt]
\multirow{2}{*}{Selection} & \multicolumn{3}{c}{Signal} & \multicolumn{2}{c}{Background} \\
\cmidrule[0.8pt](r){2-4} \cmidrule[0.8pt](l){5-6}
 & $T_{1500}$ & $T_{2000}$ & $T_{2500}$ & $\nu tb$ & $\nu Wj$ \\
\midrule[0.8pt]
No cuts & 100,000 & 100,000 & 100,000 & 500,000 & 500,000 \\
\hline
Basic selection & 87,612 (0.88) & 88,042 (0.88) & 88,305 (0.88) & 282,463 (0.56) & 312,960 (0.63) \\
\hline
Cut 1 & 29,585 (0.30) & 32,835 (0.33) & 34,167 (0.34) & 1,033 (0.002) & 13,908 (0.028) \\
\hline
Cut 2 & 20,702 (0.21) & 23,674 (0.24) & 24,910 (0.25) & 196 (3.9$\times10^{-4}$) & 57 (1.1$\times10^{-4}$) \\
\hline
Cut 3 & 14,385 (0.14) & 18,174 (0.18) & 20,308 (0.20) & 33 (6.6$\times10^{-5}$) & 18 (3.6$\times10^{-5}$) \\
\hline
Cut 4 & 5,903 (0.06) & 14,268 (0.14) & 18,325 (0.18) & 3 (6$\times10^{-6}$) & 5 (1.0$\times10^{-5}$) \\
\bottomrule[1.5pt]
\end{tabular}
\end{table}
\subsection{Analysis at $\sqrt{s}=5.29$~TeV: Hadronic $W$ channel}\label{eff:had}
\begin{figure*}[htb]
\begin{center}
\centerline{\hspace{2.0cm}\epsfxsize=8cm\epsffile{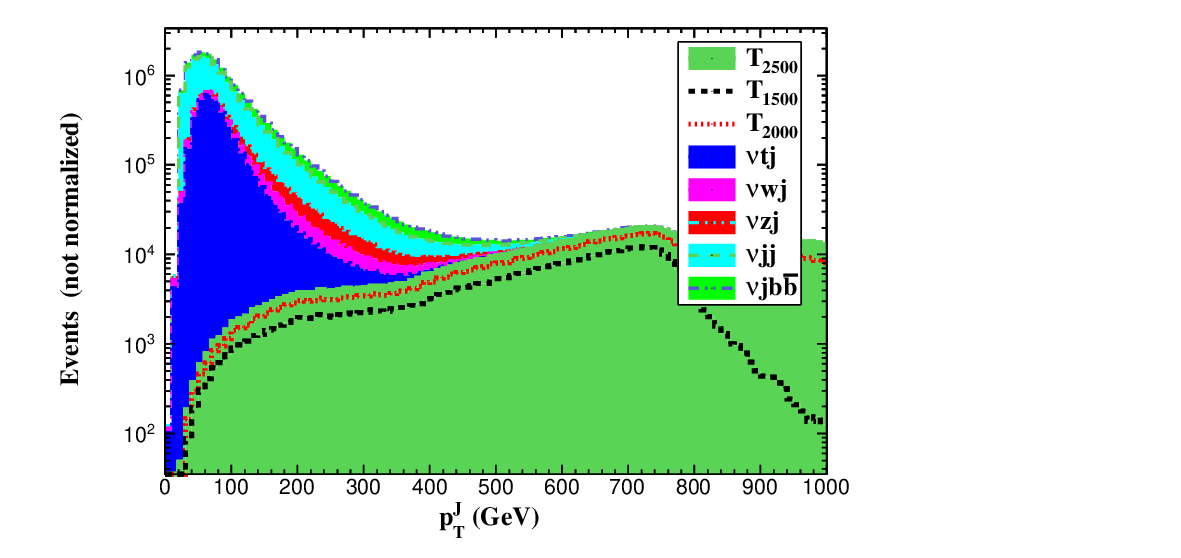}
\hspace{-2.0cm}\epsfxsize=8cm\epsffile{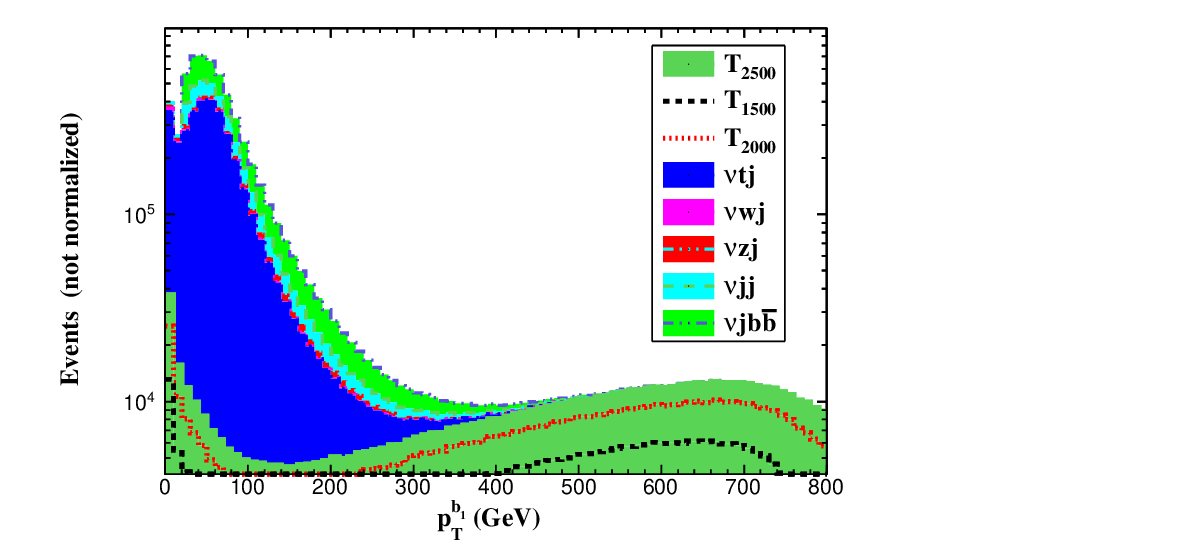}}
\centerline{\hspace{2.0cm}\epsfxsize=8cm\epsffile{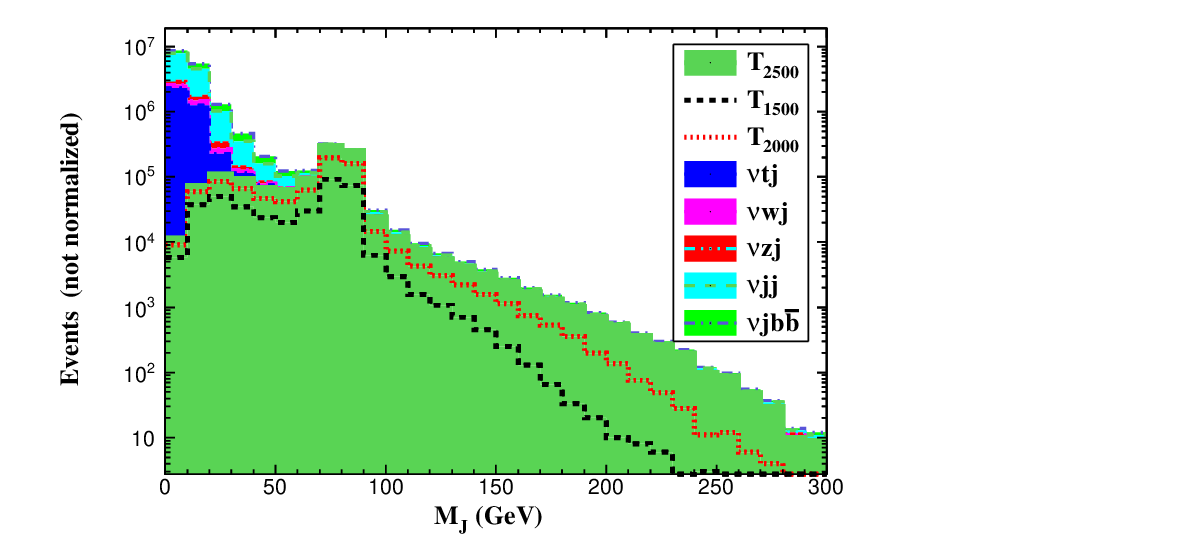}
\hspace{-2.0cm}\epsfxsize=8cm\epsffile{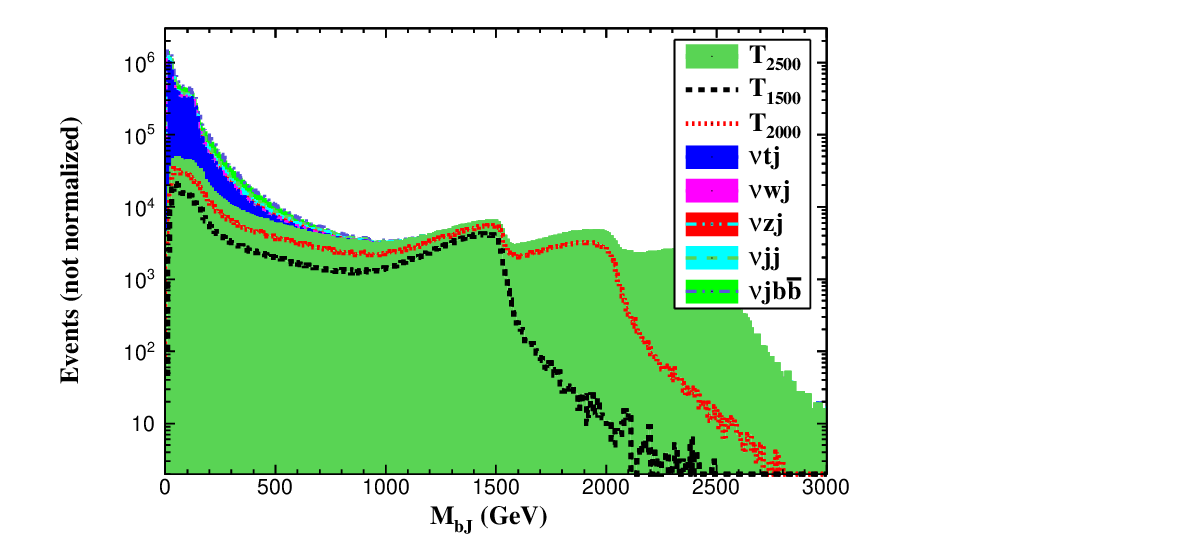}}
\caption{Preselection kinematic distributions for $p_T^{J}$, $p_T^{b_{1}}$, $M_J$, and $M_{bJ}$ in the $W_{\text{had}}$ channel at $\sqrt{s}=5.29$~TeV. The green band corresponds to $m_T = 2500$ GeV, while the dashed ($m_T = 1500$ GeV) and dotted ($m_T = 2000$ GeV) lines show independent predictions overlaid for comparison. Signal benchmarks and background contributions are shown with raw Monte Carlo statistics.}
\label{fig11}
\end{center}
\end{figure*}

Figure~\ref{fig11} displays the preselection distributions for the hadronic channel, featuring:
\begin{itemize}
    \item Transverse momentum of the fat jet ($p_T^J$)
    \item Transverse momentum of the leading $b$-jet ($p_T^{b_1}$)
    \item Invariant mass of the fat jet ($M_J$)
    \item Combined invariant mass of the $b$-jet and fat jet ($M_{bJ}$)
\end{itemize}
\end{appendix}

Table~\ref{event2} presents the complete event counts and selection efficiencies for signal and background processes in the $W_{\text{had}}$ channel at $\sqrt{s}=5.29~\text{TeV}$. The analysis includes a comprehensive simulation of the inclusive charged-current deep inelastic scattering~(DIS) background process, $\mu p \to \nu_{\mu} j b\bar{b}$, which has an initial production cross section of 439 fb after basic kinematic cuts. Upon applying the full analysis selection criteria, this background is significantly suppressed, yielding final cross sections of 0.01 fb (Cut-4a) and 0.005 fb (Cut-4b). In comparison, the total cross section of all SM backgrounds amounts to approximately 0.56 fb (Cut-4a) and 0.28 fb (Cut-4b). Given its negligible contribution relative to the total background, the DIS process can safely be disregarded.
\begin{table}[htb]
\centering
\caption{Event counts and efficiencies for the $W_{\text{had}}$ channel at $\sqrt{s}=5.29$~TeV.\label{event2}}
\scriptsize
\begin{tabular}{lrrrr}
\toprule[1.5pt]
\multirow{2}{*}{Selection} & \multicolumn{4}{c}{Signal and Background Processes} \\
\cmidrule[0.8pt](r){2-5}
 & $T_{1500}$ & $T_{2000}$ & $T_{2500}$ & $\nu tb$ \\
\midrule[0.8pt]
No cuts & 379,619 & 378,324 & 375,503 & 3,605,809 \\
\hline
Basic selection & 379,608 (1.00) & 378,314 (1.00) & 375,494 (1.00) & 3,590,580 (1.00) \\
\hline
Cut 1 & 363,168 (0.96) & 368,979 (0.98) & 369,027 (0.98) & 87,660 (0.024) \\
\hline
Cut 2 & 187,079 (0.49) & 213,026 (0.56) & 223,324 (0.59) & 2,256 (6.2$\times10^{-4}$) \\
\hline
Cut 3 & 119,531 (0.31) & 137,441 (0.36) & 144,219 (0.38) & 262 (7.3$\times10^{-5}$) \\
\hline
Cut 4a & 75,887 (0.20) & ... & ... & 10 (2.8$\times10^{-6}$) \\
\hline
Cut 4b & ... & 7,619 (0.02) & 8,838 (0.02) & 2 (5.6$\times10^{-7}$) \\
\bottomrule[1.5pt]
\end{tabular}

\vspace{0.3cm}

\begin{tabular}{lrrrr}
\toprule[1.5pt]
\multirow{2}{*}{Selection} & \multicolumn{4}{c}{Background Processes (continued)} \\
\cmidrule[0.8pt](r){2-5}
 & $\nu Wj$ & $\nu Zj$ & $\nu jj$ & $\nu jb\bar{b}$ \\
\midrule[0.8pt]
No cuts & 611,400 & 906,941 & 8,734,540 & 2,243,809 \\
\hline
Basic selection & 607,125 (0.99) & 904,826 (1.00) & 8,096,377 (0.93) & 2,198,247 (0.98) \\
\hline
Cut 1 & 85,794 (0.14) & 162,572 (0.18) & 444,173 (0.05) & 293,552 (0.13) \\
\hline
Cut 2 & 3,799 (6.2$\times10^{-3}$) & 5,850 (6.4$\times10^{-3}$) & 17,902 (2.1$\times10^{-3}$) & 15,326 (6.8$\times10^{-3}$) \\
\hline
Cut 3 & 18 (2.9$\times10^{-5}$) & 89 (9.8$\times10^{-5}$) & 52 (6$\times10^{-6}$) & 403 (1.8$\times10^{-4}$) \\
\hline
Cut 4a & 3 (4.9$\times10^{-6}$) & 31 (3.4$\times10^{-5}$) & 5 (5.7$\times10^{-7}$) & 55 (2.4$\times10^{-5}$) \\
\hline
Cut 4b & 3 (4.9$\times10^{-6}$) & 21 (2.3$\times10^{-5}$) & 2 (2.3$\times10^{-7}$) & 24 (1.1$\times10^{-5}$) \\
\bottomrule[1.5pt]
\end{tabular}

\vspace{0.2cm}
\footnotesize
\textit{Note:} Table divided for clarity. Values in parentheses show efficiencies. "..." indicates nonapplicable entries.
\end{table}


\end{document}